\documentclass[review]{elsarticle}

\usepackage{graphicx}
\usepackage{dcolumn}
\usepackage{bm}
\usepackage{braket}
\usepackage{graphicx}
\usepackage{epstopdf}
\usepackage{mwe}    
\usepackage{subfig}
\usepackage[T1]{fontenc}
\usepackage{caption} 
\usepackage{amssymb}
\usepackage{amsmath}
\usepackage{booktabs}
\usepackage{tabularx}
\usepackage{multirow}
\usepackage{float}
\usepackage{hyperref}

\newcommand{\tensr}[1]{\bm{\mathsf{#1}}} 

\begin{document}

\begin{frontmatter}
\title{Central Moment Lattice Boltzmann Method  using a Pressure-based Formulation for Multiphase Flows at High Density
Ratios and including Effects of Surface Tension and Marangoni Stresses}

\author{Farzaneh  Hajabdollahi}
\ead{farzaneh.hajabdollahiouderji@ucdenver.edu}

\author{Kannan N. Premnath}
\ead{kannan.premnath@ucdenver.edu}

\author{Samuel W. J. Welch}
\ead{Sam.Welch@ucdenver.edu}

\address{Department of Mechanical Engineering, University of Colorado Denver, 1200 Larimer street, Colorado, 80217 , U.S.A}

\date{\today}

\begin{abstract}
Simulation of multiphase flows, which are ubiquitous in nature and engineering applications, require coupled capturing or tracking of the interfaces in conjunction with the solution of fluid motion often occurring at multiple scales. In this contribution, we will present unified cascaded LB methods based on central moments for the solution of the incompressible two-phase flows at high density ratios and for capturing of the interfacial dynamics. Based on a modified continuous Boltzmann equation (MCBE) for two-phase flows, where a kinetic transformation to the distribution function involving the pressure field is introduced to reduce the associated numerical stiffness at high density gradients, a central moment cascaded LB formulation using multiple relaxation times for computing the fluid motion will be constructed. In this LB scheme, the collision step is prescribed by the relaxation of various central moments to their equilibria that are reformulated in terms of the pressure field obtained via matching to the continuous equilibria based on the transformed Maxwell distribution. Furthermore, the differential treatments for the effects of the source term representing the change due to the pressure field and of the source term due to the interfacial tension force and body forces appearing in the MCBE on different moments are consistently accounted for in this cascaded LB solver that computes the pressure and velocity fields. In addition, another cascaded LB scheme will be developed to solve for the interfacial dynamics represented by a phase field model based on the conservative Allen-Cahn equation that evolves interfaces by advection and under the competing effects due to a diffusion term and a phase segregation flux term. The latter is introduced into the cascaded LB scheme via a modification to the moment equilibria. Based on numerical simulations of a variety of two-phase flow benchmark problems at high density ratios and involving the effects of surface tension and its tangential gradients (Marangoni stresses), we will validate our unified cascaded LB approach and also demonstrate improvements in numerical stability.
\end{abstract}


\end{frontmatter}

\section{\label{sec:Intro}Introduction}
Multiphase flows arise in a number of technological and scientific applications, including in chemical and petroleum processing and power generation systems as well as microfluidic devices, and are common in nature. Such flows, whose prototypical configuration involves a continuous fluid phase and a dispersed phase, such as bubbles or droplets, are characterized by surface tension along interfaces and phase segregation effects~\cite{schwarzkopf2011multiphase}. Simulation of multiphase flows is challenging due to the simultaneous capturing or tracking of interfacial motion and the computation of fluid motion, which is generally nonlinear and can occur at multiple scales. There are various interface capturing approaches that are used in conjunction with the direct discretization of the Navier-Stokes equations (NSE), which include the volume-of-fluid method~\cite{scardovelli1999direct}, front tracking method~\cite{tryggvason2001front} and the level set method~\cite{osher2006level}.

During the last two decades, the lattice Boltzmann (LB) methods based on kinetic formulations that represent the evolution of particle distribution functions have emerged as a promising addition to the techniques available for computational fluid dynamics~\cite{he1997theory,Chen1998,Succi2001,Aidun2010,Kruger2016}. Significant interest in such methods are largely due to the locality of their
the stream-and-collide algorithm and ease of implementation of boundary conditions based on kinetic approaches on Cartesian grids. For simulation of
multiphase flows, the LB methods have been further extended to incorporate various models and techniques to represent interfacial dynamics and fluid
motion. Among them, some of the early approaches represented the phase segregation and the effect of surface tension via either a color model~\cite{gunstensen1991lattice,grunau1993lattice}, a pseudopotential formulation~\cite{shan1993lattice} or a free-energy based formulation~\cite{swift1996lattice} and their thermodynamic consistency were analyzed in~\cite{Luo2000,He2002,wagner2006thermodynamic}. A significantly improved LB method using a kinetic theory based mean field model was presented in~\cite{he1999lattice}, which allowed accurate simulation of multiphase flows at moderate density ratios. This approach used one LB scheme for the fluid motion and captured the interfacial motion via an index function, whose evolution was represented by another LB scheme where the phase segregation was achieved using a Carnahan-Starling nonideal equation of state. This was further improved for simulation of two-phase flows at high density ratios by means of a stable discretization scheme~\cite{lee2005stable}. The latter work motivated developments of consistent LB techniques for interfacial capturing techniques based on phase field models.

Phase field models represent interfaces to be diffuse, which comprise thin transitional regions of nonzero thickness across which various fluid properties vary continuously from one phase to the other~\cite{anderson1998diffuse,jacqmin1999calculation,ding2007diffuse,kim2012phase}. Such diffuse interface methods capture interfacial motion implicitly via the evolution of an order parameter, which serves as a phase field to distinguish between different fluid phases. The dynamics of the order parameter is often based on a thermodynamic free energy functional formulation, of which the Cahn-Hilliard equation (CHE)~\cite{cahn1958free} is a common choice. A LB scheme to represent the convective CHE was presented in~\cite{zheng2006lattice}, which was shown to be applicable only for density-matched two-fluid systems in~\cite{fakhari2010phase}, who then proposed a modification to handle multiphase flows at moderate density ratios. The latter work was further improved in the investigations presented in~\cite{zu2013phase,liang2014phase} to represent incompressible multiphase flows based on modified CHE for capturing of interfaces.

The challenges associated with the use of CHE, such as the need to calculate fourth order derivatives, motivated other phase field type approaches. The Allen-Cahn equation (ACE) is another type of diffuse interface model used that was originally developed for material science applications~\cite{allen1976mechanisms}. More recently, the ACE was reformulated based on a counter term approach~\cite{folch1999phase} to eliminate
curvature driven interfacial motion in order to make it applicable for two-phase flows~\cite{sun2007sharp}, in which the geometric information such
as the interface normal and curvature are computed readily by expressing them in terms of a hyperbolic tangent variation of the order parameter
across the interface. Then, Ref.~\cite{chiu2011conservative} further modified the ACE to make it mass conservative, which was shown to be equivalent
to a conservative level set approach~\cite{olsson2005conservative}. Such a conservative ACE results in a simpler formulation with less numerical
dispersion than the modified CHE, as the former requires the computation of only lower, i.e., second, order derivatives of the phase field variable when compared to the latter as noted above. Based on such conservative ACE, LB schemes for interface capturing were developed in~\cite{geier2015conservative,ren2016improved}.

The collision step plays an important role in the LB method especially for the solution of the fluid motion. The single relaxation time (SRT) model to represent the change in the distribution functions due to collision is a common approach~\cite{Qian1992}. However, it is known to be susceptible to numerical instability issues at relatively low values of the transport coefficients or at higher Reynolds number. This can be overcome to a significant extent by considering the relaxation of various raw moments to their equilibria using multiple relaxation times (MRT) to represent the effect of collisions~\cite{dHumieres2002}. A further improvement can be achieved by considering the relaxation in terms of central moments~\cite{geier2006cascaded}. It naturally maintains the Galilean invariance of all independent moments supported by a chosen lattice and the resulting method was termed as the cascaded LB method. The method was interpreted by considering relaxation in terms of a generalized equilibrium in a rest frame of reference~\cite{Asinari2008}. A scheme based on central moments to incorporate local forces and its consistency to the Navier-Stokes equations (NSE) via a Chapman-Enskog analysis was presented in~\cite{premnath2009incorporating}. Significant improvements in the numerical stability of the cascaded LB method were shown in~\cite{Geier2015,ning2016numerical}. More recently, various refinements and extensions of the central moments based LB formulation were considered (see e.g.,~\cite{lycett2016cascaded,de2017non,fei2018cascaded,hajabdollahi2018sym,elseid2018cascaded,fei2018modeling,hajabdollahi2018central,safari2018lattice,hajabdollahi2019cascaded}).

In this contribution, we present new unified cascaded LB methods for incompressible two-phase flows at high density ratios. In our formulation, one cascaded LB scheme for the solution of the multiphase fluid motion and another cascaded LB scheme for the representation of interface capturing will be developed. For the former case, the starting point is the modified continuous Boltzmann equation (MCBE) for incompressible two-phase flows~\cite{he1999lattice}, where a transformation to the distribution function is introduced to reduce the numerical stiffness associated with high density gradients and the resulting hydrodynamic variables are given in terms of the pressure and velocity fields via their zeroth and first moments, respectively. Based on this MCBE, a new discrete cascaded LB method based on central moments and multiple relaxation times for two-phase fluid flow will be constructed~\cite{hajabdollahiAPSDFD2018,Hajabdollahiphdthesis}. In this regard, we will formulate its collision step in terms of relaxation to various central moment equilibria which will be expressed by matching the central moments of the modified continuous Maxwell distribution and given in terms of the pressure field arising via the transformation mentioned above. The MCBE~\cite{he1999lattice} also contains source terms related to the pressure changes and those due to the interfacial (surface tension) force and a body force, whose respective effects on the changes in various moments are different. In order to account for the differential effects of the source term due to pressure and that due to the interfacial and body forces for handling the simulation of two-phase flows, we will present a consistent source/force treatment scheme, which is an extension of and modification to the central moment based approach that was given in a previous work for single phase flows~\cite{premnath2009incorporating}. Interfacial dynamics will be captured using the conservative ACE phase field formulation that evolves interfaces via advection and under the competing effects of a diffusion term and an interface sharpening term. In this regard, by extending the work of Ref.~\cite{geier2015conservative}, another MRT based modified cascaded LB scheme developed for the solution of the convection diffusion equation~\cite{hajabdollahi2018sym,hajabdollahi2018central}, where the sharpening term due to the phase separation flux is introduced as a modification to the moment equilibria, will be constructed to represent the evolution of the phase field variable. All fluid properties such as the density and viscosities across the phase interfaces are then expressed as smooth affine functions of the phase field variable. Since the resulting cascaded LB solvers are based on prescribing collision and sources via matching their continuous values in a moving of reference based on local fluid velocity, it naturally maintains their Galilean invariance for the independent moments supported by the chosen lattice. This can improve numerical stability for the simulation of two-phase flows at high density ratios and at relatively low fluid viscosities, thereby widening the parametric ranges for simulations. In this work, the cascaded central moment LB formulation for the coupled solution of the two-phase flow and interfacial motion will be presented on two-dimensional, nine velocity (D2Q9) lattice sets. It will then be validated for a set of numerical benchmark problems involving two-phase flows at high density ratios and including surface tension effects which are extended account for Marangoni stresses to demonstrate its accuracy and improvements
in stability.

This paper is organized as follows. In Sec.~\ref{sec:governingeqns}, we will present the governing equations for the incompressible two-phase flow and the phase field model based on the conservative ACE for the capturing of interfaces. Section~\ref{sec:DVBETwoPhaseFlows} discusses the discrete velocity Boltzmann equation for two-phase flows that represents the starting point for the construction of the central moments based kinetic formulation for its solution procedure. Then, the cascaded LB method for the solution of the two-phase flow in terms of the pressure and velocity fields is derived in Sec.~\ref{sec:cascadedLBMfluidmotion}. Subsequently, Sec.~\ref{sec:cascadedLBMinterfacialmotion} presents another cascaded LB method for interfacial dynamics based on the conservative ACE. Section~\ref{sec:resultsdiscussion} discusses the numerical validation study of the new cascaded LB formulation for a variety of two-phase flow benchmark problems, with high contrasts in fluid properties and effects of surface tension and its tangential gradients. In particular, the modeling and simulation of the effects of Marangoni stresses are discussed in Sec.~\ref{sec:resultMarangonistressdropmigration}. A comparative study of the numerical stability of different collision models in reaching low viscosities in a two-fluid system is presented in Sec.~\ref{sec:stabilitycomparison}. Finally, the conclusions of this work are summarized in Sec.~\ref{sec:summaryandconclusions}.

\section{Governing Macroscopic Equations: Interface Capturing and Two-Phase Fluid Motion\label{sec:governingeqns}}
In order to capture interfacial dynamics, we consider a phase field method based on the conservative Allen-Cahn equation (ACE). This was originally formulated for two-phase flows by removing the curvature-driven motion~\cite{sun2007sharp} via a counter term approach~\cite{folch1999phase} and then re-expressed in a conservative form~\cite{chiu2011conservative}. Let $\phi$ be an order parameter or the phase field variable, with $\phi=\phi_A$ representing the fluid in phase $A$ and $\phi=\phi_B$ denoting that in phase $B$. Then, the interface propagation given in terms of the phase field variable based on the conservative ACE can be written as
\begin{equation}
\frac{\partial \phi}{\partial t}+\bm{\nabla}\cdot(\phi\bm{u})=\bm{\nabla}\cdot\left[M_\phi(\bm{\nabla}\phi-\theta\bm{n})\right],
\label{eq:ACE}
\end{equation}
where $\bm{u}$ is the fluid velocity, $\bm{n}$ is the unit normal vector, which can be computed via the order parameter $\phi$ as $\bm{n}=\frac{\bm{\nabla} \phi}{|\bm{\nabla} \phi|}$, and $M_\phi$ is the mobility. In the above, the variable $\theta$ can be expressed as
\begin{equation}
\theta=\frac{-4(\phi-\phi_A)(\phi-\phi_B)}{W (\phi_A-\phi_B)},\label{eq:thetavariable}
\end{equation}
where the parameter $W$ is related to the width of the interface. The right hand side of Eq.~(\ref{eq:ACE}) is obtained by removing the curvature-driven
interface motion $u_\kappa\bm{n}=-M_\phi\kappa_m\bm{n}$ by canceling it out by adding a counteracting term based on computing the curvature $\kappa_m$, where $\kappa_m=\bm{\nabla}\cdot\bm{n}$ with $\bm{n}=\frac{\bm{\nabla} \phi}{|\bm{\nabla} \phi|}$, directly in terms of a kernel function given by the following hyperbolic tangent profile of the order parameter
\begin{equation}
\phi(\zeta)=\frac{1}{2}(\phi_A+\phi_B)+\frac{1}{2}(\phi_A-\phi_B)\tanh\left(\frac{2\zeta}{W}\right),\label{eq:hyperbolictangentprofile}
\end{equation}
which represents the equilibrium profile of the phase field variable, where $\zeta$ is a spatial coordinate along the normal with the origin at the interface. Thus, Eq.~(\ref{eq:ACE}) effectively represents the relaxation of any arbitrary initial distribution of the order parameter to a hyperbolic tangent profile across the interface, which is then sustained during interfacial advection. Equivalently, this equation can be interpreted as the interface propagating via advection (given by its LHS) under the competing effects of a diffusion term and an interface sharpening term or a separation flux term (given by the first and second terms on the RHS, respectively). In the above, $W$ and $M_\phi$ are numerical parameters, with $W$ representing the interface thickness, while $M_\phi$ controlling the relaxation rate of any initial
$\phi$ to its equilibrium profile across the interfaces (Eq.~(\ref{eq:hyperbolictangentprofile})) as well as the dissipation of any interface singularities via diffusion.

On the other hand, the two-phase fluid flow is represented by the following incompressible Navier-Stokes equations (NSE):
\begin{eqnarray}
\bm{\nabla}\cdot \bm{u}&=&0,\label{eq:incompressibleNSE1}\\
\rho\left(\frac{\partial \bm{u}}{\partial t}+\bm{\nabla}\cdot(\bm{u}\bm{u})\right)&=&-\bm{\nabla}p+\bm{\nabla}\cdot\left[\mu(\bm{\nabla}\bm{u}+\bm{\nabla}\bm{u}^\dag)\right]+\bm{F}_s+\bm{F}_{ext},
\label{eq:incompressibleNSE2}
\end{eqnarray}
where $p$ is the hydrodynamic pressure, $\rho$ is the fluid density, $\mu$ is its viscosity, $\bm{F}_s$ is the smoothed formulation of the surface tension force and $\bm{F}_{ext}$ is an external body force (e.g., gravity).

In the above, there are several ways to express the surface tension force $\bm{F}_s$ as a smoothed representation based on the order parameter. One approach is based on a thermodynamic (Gibbs-Duhem) formulation in which the surface tension force is calculated from the negative product of the gradient of the chemical potential $\tilde{\mu}_\phi$ and the phase field variable $\phi$ as follows (see e.g.,~\cite{jacqmin1999calculation}):
\begin{equation}
\bm{F}_s=-\phi\bm{\nabla}\tilde{\mu}_\phi, \quad \tilde{\mu}_\phi=4 \beta(\phi-\phi_A)(\phi-\phi_B)\left(\phi-\left(\phi_A+\phi_B\right)/2\right)-\kappa \bm{\nabla}^2\phi.\label{eq:surfacetensionthermodynamic}
\end{equation}
Here, the parameters $\beta$ and $\kappa$ are used to control the surface tension $\sigma$ and the interface thickness $W$ via the following relations
\begin{equation}
\kappa=\frac{3}{2}\sigma W,\quad \beta=\frac{12\sigma}{W}.
\end{equation}
Alternatively, geometric approaches such as the continuous surface force formulation can be considered~\cite{popinet2018numerical}. In particular, a geometric approach for the surface tension force developed originally for level set methods and adapted for phase field methods~\cite{kim2005continuous} can be written as
\begin{equation}
\bm{F}_s=-\tilde{\kappa}|\bm{\nabla} \phi|^2\left(\bm{\nabla}\cdot\bm{n}\right)\bm{n}.\label{eq:surfacetensiongeometric}
\end{equation}
Here, the parameter $\tilde{\kappa}$ is related to the surface tension $\sigma$ via $\tilde{\kappa}=\gamma\sigma W$, where the coefficient $\gamma$ satisfies $\gamma\int_{-\infty}^{\infty}(d\phi/d\zeta)^2d\zeta=1$, which arises from interpreting the surface tension in terms of interfacial energy per unit surface area by considering the equilibrium phase field variable profile given in Eq.~(\ref{eq:hyperbolictangentprofile}) and matching it with the sharp interface limit for a flat interface~\cite{kim2005continuous}. In this work, this latter (geometric) approach is adopted for representing the surface tension force $\bm{F}_s$ for performing two-phase flow simulations using cascaded LB formulations discussed in what follows. Finally, the jumps in fluid properties such as the density and viscosity across the interface are smoothed as well and can be written as a continuous function of the phase field variable $\phi$ and then used in Eq.~(\ref{eq:incompressibleNSE2}) in different ways. In this study, we employ a linear interpolation for representing the interfacial variations of the fluid properties (see e.g.,~\cite{ding2007diffuse}). Thus,
\begin{equation}
\rho=\rho_B+\frac{\phi-\phi_A}{\phi_A-\phi_B}(\rho_A-\rho_B),\quad
\mu=\mu_B+\frac{\phi-\phi_A}{\phi_A-\phi_B}(\mu_A-\mu_B),\label{eq:sfluidpropertiesinterpolation}
\end{equation}
where $\rho_A$ and $\rho_B$ are the densities and $\mu_A$ and $\mu_B$ are the dynamic viscosities in the fluid phases denoted by $\phi_A$ and $\phi_B$, respectively. In this work, we consider $\phi_B=0$ and $\phi_A=1$.

\section{Modified Continuous Boltzmann Equation for Two-Phase Flows and Central Moments of Equilibria and Sources\label{sec:DVBETwoPhaseFlows}}
To solve the incompressible Navier-Stokes equations (NSE) for two-phase flows (Eqs.~(\ref{eq:incompressibleNSE1}) and (\ref{eq:incompressibleNSE2})) in a kinetic formulation, the starting point is the two-dimensional (2D) continuous Boltzmann equation given by~\cite{he1999lattice}
\begin{equation}
\frac{Df}{Dt}\equiv \frac{\partial f}{\partial t}+\bm{\xi}\cdot\bm{\nabla}f=-\frac{1}{\tau}\left(f-f^M\right)+\frac{\left(\bm{\xi}-\bm{u}\right)}{\rho c_s^2}\cdot\left(\bm{F}_t-\bm{\nabla}\psi\right)f^M,
\label{eq:CBE}
\end{equation}
where $f=f(\bm{x},t; \bm{\xi})$  is the density distribution function at a location $\bm{x}$ and at time $t$, corresponding to the particle velocity $\bm{\xi}=(\xi_x,\xi_y)$. Here, $ f^M$ is the local Maxwell distribution function defined as
\begin{equation}
f^M\equiv f^M(\rho,\bm{u})=\frac{\rho}{2\pi c_s^2}\exp\left[-\frac{(\bm{\xi}-\bm{u})^2}{2c_s^2}\right],
\label{eq:Maxwelldistribution}
\end{equation}
where $c_s$ is the speed of sound and fluid velocity $\bm{u}=(u_x,u_y)$. The effect of collisions is typically represented as a relaxation of $f$ to its equilibrium, i.e., $f^M$ with a characteristic time scale $\tau$. The continuous formulation of the interfacial tension force $\bm{F}_s$, which is discussed in the previous section, along with any local body force $\bm{F}_{ext}$ are grouped as the total force $\bm{F}_t=\bm{F}_s+\bm{F}_{ext}$. This total force along with the gradient contribution of the net effect of the hydrodynamic pressure $p$ relative to that from the ideal equation of state $\rho c_s^2$, i.e., $\psi(\rho)=p-\rho c_s^2$ are accounted for via a source term in Eq.~(\ref{eq:CBE}). In general, multiphase flows can be associated with relatively large jumps in fluid properties across the interfaces. In particular, as the density gradients $\bm{\nabla}\rho$ or $\bm{\nabla}\psi$ become relatively large, Eq.~(\ref{eq:CBE}) becomes numerically stiff.

To alleviate such numerical stiffness, the following kinetic transformation to the distribution can be introduced~\cite{he1999lattice}
\begin{equation}
g=fc_s^2+(p-\rho c_s^2)\frac{f^M}{\rho}(\rho,\bm{0}),\label{eq:kinetictransformation}
\end{equation}
where $g$ can be regarded as the pressure distribution function. Here, $f^M(\rho,\bm{0})$ is the local Maxwellian with null macroscopic fluid velocity, which follows from Eq.~(\ref{eq:Maxwelldistribution}) as
\begin{equation}
f^M(\rho,\bm{0})=\frac{\rho}{2\pi c_s^2}\exp\left[-\frac{\bm{\xi}^2}{2c_s^2}\right].
\end{equation}
Then, by applying the above transformation (Eq.~(\ref{eq:kinetictransformation})) to the continuous Boltzmann equation (Eq.~(\ref{eq:CBE})) and assuming two-phase flows in the incompressible limit, i.e., $|\bm{u}|\ll 1$, the following kinetic equation for the distribution function $g$ can be obtained~\cite{he1999lattice}
\begin{eqnarray}
\frac{Dg}{Dt}=&&-\frac{1}{\tau}\left(g-g^{eq}\right)+\left(\bm{\xi}-\bm{u}\right)\cdot\bm{F}_t\frac{f^M(\rho,\bm{u})}{\rho}+\nonumber\\
&&\left(\bm{\xi}-\bm{u}\right)\cdot\bm{F}_p\underbrace{\left\{\frac{f^M(\rho,\bm{u})}{\rho}-\frac{f^M(\rho,\bm{0})}{\rho}\right\}}_{O(\bm{u})},
\label{eq:MCBE}
\end{eqnarray}
which is referred to as the modified continuous Boltzmann equation (MCBE) in this work. In Eq.~(\ref{eq:MCBE}), $g^{eq}$ is the transformed local Maxwellian or the modified equilibrium distribution function, which reads as
\begin{equation}
g^{eq}=c_s^2f^M(\rho,\bm{u})+(p-\rho c_s^2)\frac{f^M}{\rho}(\rho,\bm{0})
\label{eq:modifiedeqm}
\end{equation}
and $\bm{F}_p$ is the net effect of the hydrodynamic pressure $p$ relative to the contribution from the ideal equation of state dependent on density, which is referred as the net gradient pressure force, and can be expressed as
\begin{equation}
\bm{F}_p=-\bm{\nabla}(p-\rho c_s^2)\equiv-\bm{\nabla}\psi. \label{eq:netgradientpressureforce}
\end{equation}
In the MCBE (Eq.~(\ref{eq:MCBE})), even though $\bm{F}_p$ can be large at high density ratios, since it is multiplied by ${\left\{\frac{f^M(\rho,\bm{u})}{\rho}-\frac{f^M(\rho,\bm{0})}{\rho}\right\}}$, which is $O(\bm{u})$ and small, the associated numerical stiffness issues on the evolution of the distribution function $g$ is reduced significantly. Hence, the MCBE serves as the starting point in the construction of a discrete kinetic scheme for the solution of the incompressible two-phase flows with high phase density contrasts. The hydrodynamic pressure and velocity fields are then obtained as the zeroth and first kinetic moments of the distribution function $g$, respectively. That is,
\begin{equation}
p=\int_{-\infty}^{\infty}\int_{-\infty}^{\infty}g d\xi_xd\xi_y,\quad \rho c_s^2\bm{u}=\int_{-\infty}^{\infty}\int_{-\infty}^{\infty}g\bm{\xi} d\xi_xd\xi_y.\label{eq:gkineticmoments}
\end{equation}
\subsection{Continuous Central Moments of Equilibria and Sources of MCBE}
As a prelude to constructing a cascaded LB scheme from the discretization of the MCBE, which is discussed in the next section, we will first need the continuous central moments of its equilibria and various sources. They are based on the contributions from the corresponding continuous Maxwell distribution function evaluated with and without the macroscopic fluid velocity in view of the kinetic transformation introduced above.

First, defining the continuous central moments of the local Maxwellian for a moving fluid, i.e., with the macroscopic fluid velocity, of order $(m+n)$ as
\begin{equation}
\hat{\Pi}_{mn}^{M}=\int_{-\infty}^{\infty}\int_{-\infty}^{\infty}f^M(\rho,\bm{u})(\xi_x-u_x)^m(\xi_y-u_y)^n d\xi_xd\xi_y,
\label{eq:centralmomentsMaxwellian}
\end{equation}
and then defining the continuous central moments of the local Maxwellian with the null macroscopic fluid velocity of order $(m+n)$ as
\begin{equation}
\hat{\Pi}_{mn}^{M(\bm{0})}=\int_{-\infty}^{\infty}\int_{-\infty}^{\infty}f^M(\rho,\bm{0})(\xi_x-u_x)^m(\xi_y-u_y)^n d\xi_xd\xi_y.
\label{eq:centralmomentsMaxwelliannull}
\end{equation}
The definite integrals given in Eqs.~(\ref{eq:centralmomentsMaxwellian}) and (\ref{eq:centralmomentsMaxwelliannull}) can be evaluated exactly via standard quadrature rules. The D2Q9 lattice used in the construction of the cascaded LB scheme based on a matching principle in the next section supports nine independent moment components. In this regard, we will need the corresponding components of $\hat{\Pi}_{mn}^{M}$ and $\hat{\Pi}_{mn}^{M(\bm{0})}$ as intermediate results. Thus, calculating the components of the continuous central moments of the Maxwellian $\hat{\Pi}_{mn}^{M}$ (Eq.~(\ref{eq:centralmomentsMaxwellian})) at various orders, which read as
\begin{eqnarray}
&&\hat{\Pi}_{00}^{M}=\rho,\quad \hat{\Pi}_{10}^{M}=0,\quad \hat{\Pi}_{01}^{M}=0,\quad \hat{\Pi}_{20}^{M}=\rho c_s^2,\quad \hat{\Pi}_{02}^{M}=\rho c_s^2,\quad \hat{\Pi}_{11}^{M}=0,\nonumber\\
&&\hat{\Pi}_{21}^{M}=0,\quad \hat{\Pi}_{12}^{M}=0,\quad \hat{\Pi}_{22}^{M}=\rho c_s^4.\label{eq:centralmomentsMaxwelliannulcomponents}
\end{eqnarray}
and those of $\hat{\Pi}_{mn}^{M(\bm{0})}$ (Eq.~(\ref{eq:centralmomentsMaxwelliannull})) may be written as
\begin{eqnarray}
&&\hat{\Pi}_{00}^{M(\bm{0})}=1,\quad \hat{\Pi}_{10}^{M(\bm{0})}=-u_x,\quad \hat{\Pi}_{01}^{M(\bm{0})}=-u_y,\quad \hat{\Pi}_{20}^{M(\bm{0})}=(u_x^2+c_s^2),\quad \hat{\Pi}_{02}^{M(\bm{0})}=(u_y^2+c_s^2),\nonumber\\
&&\hat{\Pi}_{11}^{M(\bm{0})}=u_xu_y,\quad
\hat{\Pi}_{21}^{M(\bm{0})}=-(u_x^2+c_s^2)u_y,\quad \hat{\Pi}_{12}^{M(\bm{0})}=-(u_y^2+c_s^2)u_x,\quad \hat{\Pi}_{22}^{M(\bm{0})}=(u_x^2+c_s^2)(u_y^2+c_s^2).\nonumber\label{eq:centralmomentsMaxwelliannulcomponentsnull}
\end{eqnarray}
Then, in order to discretize Eq.~(\ref{eq:MCBE}) in a cascaded LB formulation, we need the continuous central moments of the equilibrium pressure distribution function or the transformed Maxwellian $g^{eq}$ (Eq.~(\ref{eq:modifiedeqm})) of order $(m+n)$. By defining it as
\begin{equation}
\hat{\Pi}_{mn}^{eq,g}=\int_{-\infty}^{\infty}\int_{-\infty}^{\infty}g^{eq}(\xi_x-u_x)^m(\xi_y-u_y)^n d\xi_xd\xi_y,
\label{eq:centralmomentsgeq}
\end{equation}
it readily follows that Eq.~(\ref{eq:centralmomentsgeq}) satisfies the following relation
\begin{equation*}
\hat{\Pi}_{mn}^{eq,g}=c_s^2\hat{\Pi}_{mn}^{M}+\psi(\rho)\hat{\Pi}_{mn}^{M(\bm{0})}.
\end{equation*}
Evaluating its nine components, we get
\begin{eqnarray}
&&\hat{\Pi}_{00}^{eq,g}=p,\quad \hat{\Pi}_{10}^{eq,g}=-\psi(\rho)u_x,\quad \hat{\Pi}_{01}^{eq,g}=-\psi(\rho)u_y,\quad \hat{\Pi}_{20}^{eq,g}=p c_s^2+\psi(\rho)u_x^2,\nonumber\\
&& \hat{\Pi}_{02}^{eq,g}=p c_s^2+\psi(\rho)u_y^2, \quad \hat{\Pi}_{11}^{eq,g}=\psi(\rho)u_xu_y, \quad \hat{\Pi}_{21}^{eq,g}=-\psi(\rho)(c_s^2+u_x^2)u_y,\nonumber\\
&&\quad \hat{\Pi}_{12}^{eq,g}=-\psi(\rho)(c_s^2+u_y^2)u_x,\quad\hat{\Pi}_{22}^{eq,g}=c_s^6\rho+\psi(\rho)(u_x^2+c_s^2)(u_y^2+c_s^2).\label{eq:centralmomentsgeqcomponents}
\end{eqnarray}

Next, we need the continuous central moments of the source term due to the total (interfacial and local body) force $\bm{F}_t=(F_{tx},F_{ty})$ of order $(m+n)$ in MCBE (Eq.~(\ref{eq:MCBE})), which can be defined as
\begin{equation}
\hat{\Gamma}_{mn}^{t}=\int_{-\infty}^{\infty}\int_{-\infty}^{\infty}S^{t}(\xi_x-u_x)^m(\xi_y-u_y)^n d\xi_xd\xi_y,
\label{eq:centralmomentstotalforce}
\end{equation}
where
\begin{equation}
S^t=\left(\bm{\xi}-\bm{u}\right)\cdot\bm{F}_t\frac{f^M(\rho,\bm{u})}{\rho}.\label{eq:sourcetotalforce}
\end{equation}
It can be shown that this continuous central moment satisfies the following identity that depends on the those of the Maxwellian
\begin{equation*}
\hat{\Gamma}_{mn}^{t}=F_{tx}\frac{\hat{\Pi}_{m+1,n}^{M}}{\rho}+F_{ty}\frac{\hat{\Pi}_{m,n+1}^{M}}{\rho}.
\end{equation*}
By evaluating its components and dealiasing the resulting central moment components higher than the second order by setting them to zero, as they do not
influence the recovery of the hydrodynamics via the Chapman-Enskog expansion~\cite{premnath2009incorporating}, the results can be summarized as
\begin{eqnarray}
&&\hat{\Gamma}_{00}^{t}=0,\quad \hat{\Gamma}_{10}^{t}=c_s^2F_{tx},\quad \hat{\Gamma}_{01}^{t}=c_s^2F_{ty},\quad \hat{\Gamma}_{20}^{t}=0,\quad \hat{\Gamma}_{02}^{t}=0,\quad \hat{\Gamma}_{11}^{t}=0,\nonumber\\
&&\hat{\Gamma}_{21}^{t}=0,\quad \hat{\Gamma}_{12}^{t}=0,\quad \hat{\Gamma}_{22}^{t}=0.\label{eq:centralmomentstotalforcecomponents}
\end{eqnarray}
Finally, we define the continuous central moments of the source term due to the net gradient pressure force $\bm{F}_p=(F_{px},F_{py})$ in Eq.~(\ref{eq:MCBE}) of order $(m+n)$ as
\begin{equation}
\hat{\Gamma}_{mn}^{p}=\int_{-\infty}^{\infty}\int_{-\infty}^{\infty}S^{p}(\xi_x-u_x)^m(\xi_y-u_y)^n d\xi_xd\xi_y,
\label{eq:centralmomentsnetgradientpressureforce}
\end{equation}
where
\begin{equation}
S^p=\left(\bm{\xi}-\bm{u}\right)\cdot\bm{F}_p\left\{\frac{f^M(\rho,\bm{u})}{\rho}-\frac{f^M(\rho,\bm{0})}{\rho}\right\}.\label{eq:sourcenetgradientpressureforce}
\end{equation}
Based on its definition, this central moment $\hat{\Gamma}_{mn}^{p}$ can be demonstrated to satisfy the following identity
\begin{equation*}
\hat{\Gamma}_{mn}^{p}=F_{px}\left(\frac{\hat{\Pi}_{m+1,n}^{M}}{\rho}-\frac{\hat{\Pi}_{m+1,n}^{M(\bm{0})}}{\rho}\right)+F_{py}\left(\frac{\hat{\Pi}_{m,n+1}^{M}}{\rho}-\frac{\hat{\Pi}_{m,n+1}^{M(\bm{0})}}{\rho}\right).
\end{equation*}
By using this and deriving the expressions for the nine components, where, as before, we retain the results only up to the second order moments that determine the two-phase fluid motion and set the higher order ones to zero, they can be summarized as
\begin{eqnarray}
&&\hat{\Gamma}_{00}^{p}=(F_{px}u_x+F_{py}u_y),\quad \hat{\Gamma}_{10}^{p}=-u_x\hat{\Gamma}_{00}^{p},\quad \hat{\Gamma}_{01}^{p}=-u_y\hat{\Gamma}_{00}^{p}, \quad \hat{\Gamma}_{20}^{p}=2c_s^2F_{px}u_x+(u_x^2+c_s^2)\hat{\Gamma}_{00}^{p},\nonumber\\ &&\hat{\Gamma}_{02}^{p}=2c_s^2F_{py}u_y+(u_y^2+c_s^2)\hat{\Gamma}_{00}^{p},\quad \hat{\Gamma}_{11}^{p}=c_s^2(F_{px}u_y+F_{py}u_x)+u_xu_y\hat{\Gamma}_{00}^{p},\nonumber\\
&&\hat{\Gamma}_{21}^{p}=0,\quad \hat{\Gamma}_{12}^{p}=0,\quad \hat{\Gamma}_{22}^{p}=0.\label{eq:centralmomentsnetgradientpressureforcecomponents}
\end{eqnarray}

\section{Cascaded LB Method for Solution of Two-Phase Fluid Motion\label{sec:cascadedLBMfluidmotion}}
We will now present a cascaded central moment LB method based on the discretization of the MCBE discussed in the previous section for the
solution of incompressible two-phase flow. In this regard, we consider the D2Q9 lattice, whose components of the particle velocities
are represented by the following vectors using the standard Dirac's bra-ket notation:
\begin{subequations}
\begin{eqnarray}
&\ket{e_{x}} =\left(     0,     1,    0,     -1,     0,  1, -1, -1,  1  \right)^\dag,\label{eq:15a}\\
&\ket{e_{y}} =\left(     0,     0,     1,     0,    -1,  1,  1, -1, -1
\right)^\dag.\label{eq:particlevelocities}
\end{eqnarray}
\end{subequations}
In addition, we need to define the following nine-dimensional vector
\begin{eqnarray}
&\ket{1} =\left(     1,     1,    1,     1,     1,  1,  1,  1,  1  \right)^\dag, \label{eq:unitvectorbasis}
\end{eqnarray}
whose inner product with a discrete distribution function $g_{\alpha}$ (see below), where $\alpha=0,1,2,\cdots,8$ represents the particle velocity direction, i.e., its zeroth moment yields the pressure field. Using the above, the following set of orthogonal moment basis vectors can be used to construct the cascaded LB formulation:
\begin{eqnarray}
\ket{K_0}=\ket{1}, \quad \ket{K_1}=\ket{e_{x}}, \quad \ket{K_2}=\ket{e_{y}}, \quad \ket{K_3}=3\ket{e_{x}^2+e_{y}^2}-4\ket{1},\nonumber \\
\ket{K_4}=\ket{e_{x}^2-e_{y}^2}, \quad
\ket{K_5}=\ket{e_{x}e_{y}}, \quad
\ket{K_6}=-3\ket{e_{x}^2e_{y}}+2\ket{e_{y}},\nonumber \\
\ket{K_7}=-3\ket{e_{x}e_{y}^2}+2\ket{e_{x}}, \quad
\ket{K_8}=9\ket{e_{x}^2e_{y}^2}-6\ket{e_{x}^2+e_{y}^2}+4\ket{1}.\label{eq:orthogonalbasisvectors}
\end{eqnarray}
In the above, a symbol such as $\ket{e_x^2e_y}=\ket{e_xe_xe_y}$ represents a vector resulting from the elementwise vector multiplication of the sequence of vectors $\ket{e_x}$, $\ket{e_x}$ and $\ket{e_y}$. By combining the above nine independent vectors, we then obtain the following orthogonal moment basis matrix
\begin{equation}
\tensr{K}=\left[\ket{K_0},\ket{K_1},\ket{K_2},\ket{K_3},\ket{K_4},\ket{K_5},\ket{K_6},\ket{K_7},\ket{K_8}\right].
\label{eq:collisionmatrix}
\end{equation}

Then, we perform the standard spatial and temporal discretization of the MCBE (Eq.~(\ref{eq:MCBE})) along the characteristic directions of the particle velocities over a time step $\delta_t$ (typically $\delta_t=1$ in lattice units), where we apply a trapezoidal rule for the treatment of the source term to maintain a second order accuracy~\cite{he1999lattice}, which yields
\begin{equation}
g_{\alpha}(\bm{x}+\bm{e}_\alpha\delta_t,t+\delta_t)=g_{\alpha}(\bm{x},t)+(\tensr K\cdot \widehat {\mathbf h})_{\alpha}+\frac{1}{2}\left[S_\alpha(\bm{x},t)+S_\alpha(\bm{x}+\bm{e}_\alpha\delta_t,t+\delta_t)\right]\delta_t.
\label{eq:cascadedLBEtwophaseflow}
\end{equation}
Here, $(\tensr K\cdot \widehat {\mathbf h})_{\alpha}$ is the cascaded collision operator, where $\mathbf{\widehat{h}}=\ket{\widehat{h}_{\alpha}}=(\widehat{h}_0,\widehat{h}_1,\widehat{h}_2,\ldots,\widehat{h}_8)^\dag$ is a vector representing the
changes in all the nine moments supported by the lattice under collision which will be determined in what follows. $S_\alpha$ is the total source term
representing the cumulative effect of the discrete version of the source due to the interfacial and local body force $S_\alpha^t$ (via Eq.~(\ref{eq:sourcetotalforce})) and that due to the net gradient pressure force $S_\alpha^p$ (via Eq.~(\ref{eq:sourcenetgradientpressureforce})):
\begin{equation}
S_\alpha=S_\alpha^t+S_\alpha^p.\label{eq:sourcetermcumulative}
\end{equation}
In order to remove implicitness in Eq.~(\ref{eq:cascadedLBEtwophaseflow}), we apply a variable transformation $\overline{g}_\alpha=g_\alpha-\frac{1}{2}S_\alpha\delta_t$, which then results in the following effectively explicit cascaded LB scheme
\begin{equation}
\overline{g}_{\alpha}(\bm{x}+\bm{e}_\alpha\delta_t,t+\delta_t)=\overline{g}_{\alpha}(\bm{x},t)+(\tensr K\cdot \widehat {\mathbf h})_{\alpha}+\delta g_\alpha^s,
\label{eq:cascadedLBEtwophaseflowexplicit}
\end{equation}
where $\delta g_\alpha^s$ is a modified cumulative source term under the variable transformation, which we prescribe to be the following:
\begin{equation}
\delta g_\alpha^s = \tensr{K}^{-1}\left(\tensr{I}-\frac{1}{2}\hat{\Lambda}\right)\tensr{K}\mathbf{S}.\label{eq:sourcetermcumulativetransformed}
\end{equation}
Here, $\mathbf{S}=(S_0,S_1,S_2,\ldots,S_8)^\dag $ represents a vector of all the nine components of the discrete source term and $\hat{\Lambda}=\mbox{diag}(\omega_0,\omega_1,\omega_2,\ldots,\omega_8)$ is a relaxation time matrix used in the development of the cascaded collision operator under relaxation of different central moments later. Since the effects of the two sources $S_\alpha^t$ and $S_\alpha^p$ appearing in the cumulative source term $S_\alpha$ on the changes of various moments are different, we consider a modification to the earlier central
moments based strategy~\cite{premnath2009incorporating} in this regard. The expression given in Eq.~(\ref{eq:sourcetermcumulativetransformed}) is motivated to
remove any spurious effects due to the source term in the second order non-equilibrium moments, which are related to the viscous stress tensor, in order to consistently recover the incompressible NSE for two-phase flows. Similar approach has been considered in the MRT-LBE with forcing term previously (see e.g.,~\cite{Premnath2007}), but the form of $\delta g_\alpha^s$ in Eq.~(\ref{eq:sourcetermcumulativetransformed}) will be still determined by a central moments based strategy in what follows.

In order of derive the expressions for $\widehat {\mathbf h}$ and $\delta g_\alpha^s$ to complete the formulation of the cascaded LB scheme for two-phase fluid motion, we first define the discrete central moments of the distribution function, its equilibrium and the source term as
\begin{eqnarray}
\left( {\begin{array}{*{20}{l}}
{{{\hat \eta }_{mn}}}\\
{\hat {\eta} _{mn}^{eq}}\\
{\hat {\sigma} _{mn}}\\
{{{\hat{\overline{\eta}}}_{mn}}}
\end{array}} \right) = \sum\limits_\alpha  {\left( {\begin{array}{*{20}{l}}
{{g_\alpha}}\\
{g_\alpha ^{eq}}\\
{S_\alpha}\\
{{\overline{g}_\alpha}}
\end{array}} \right)} {{(e_{\alpha x}-u_x)}^m}{{(e_{\alpha y}-u_y)}^n},
\label{eq:discretecentralmoments}
\end{eqnarray}
where $\hat{\overline{\eta}}_{mn}=\hat{\eta}_{mn}-\frac{1}{2}\hat {\sigma} _{mn}\delta_t$ and the corresponding raw moments as
\begin{eqnarray}
\left( {\begin{array}{*{20}{l}}
{{{\hat \eta }_{mn}^{'}}}\\
{\hat {\eta} _{mn}^{eq '}}\\
{\hat {\sigma} _{mn}^{'}}\\
{{{\hat{\overline{\eta}}}_{mn}^{'}}}
\end{array}} \right) = \sum\limits_\alpha  {\left( {\begin{array}{*{20}{l}}
{{g_\alpha}}\\
{g_\alpha ^{eq}}\\
{S_\alpha}\\
{{\overline{g}_\alpha}}
\end{array}} \right)} {{e_{\alpha x}^m}}{{e_{\alpha y}^n}}, \label{eq:discreterawmoments}
\end{eqnarray}
where $\hat{\overline{\eta}}_{mn}^{'}=\hat{\eta}_{mn}^{'}-\frac{1}{2}\hat {\sigma} _{mn}^{'}\delta_t$. Then, we need to determine the expressions for the discrete central moments of the equilibrium distribution function and the source term. In this regard, we apply a matching principle~\cite{geier2006cascaded,premnath2009incorporating}, where they are respectively set equal to their continuous values for all orders supported by the lattice. That is,
\begin{equation}
\hat{\eta}_{mn}^{eq}=\hat{\Pi}_{mn}^{eq,g}, \quad \hat{\sigma}_{mn}=\hat{\Gamma}_{mn}\equiv \hat{\Gamma}_{mn}^t+\hat{\Gamma}_{mn}^p, \label{eq:matchingmoments}
\end{equation}
where the continuous central moment components of the equilibrium $\hat{\Pi}_{mn}^{eq,g}$ is given in Eq.~(\ref{eq:centralmomentsgeqcomponents}), while those for the source terms $\hat{\Gamma}_{mn}^t$ and $\hat{\Gamma}_{mn}^p$ can be found in Eqs.~(\ref{eq:centralmomentstotalforcecomponents}) and (\ref{eq:centralmomentsnetgradientpressureforcecomponents}), respectively. This step effectively preserves the Galilean invariance of all the
moments independently supported by the lattice.

Based on Eq.~(\ref{eq:matchingmoments}), the first step in deriving the modified cumulative source term in the velocity space due to various sources/forces $\delta g_\alpha^s$ is to convert the central moments $\hat{\sigma}_{mn}$ to the corresponding raw moments $\hat{\sigma}_{mn}^{'}$ at various orders via the binomial transform. Performing this and setting all the cumulative source moments of second and higher order to zero as they do not affect recovering the hydrodynamics of the two-phase fluids in the Chapman-Enskog analysis~\cite{premnath2009incorporating,hajabdollahi2018sym}, we get
\begin{eqnarray*}
&&\hat{\sigma}_{00}^{'}=\hat{\Gamma}_{00}^p\equiv(F_{px}u_x+F_{py}u_y), \quad \hat{\sigma}_{10}^{'}=c_s^2F_{tx},\quad \hat{\sigma}_{01}^{'}=c_s^2F_{ty},\\
&&\hat{\sigma}_{20}^{'}=2c_s^2(F_{tx}u_x+F_{px}u_x)+c_s^2\hat{\Gamma}_{00}^p,\quad \hat{\sigma}_{02}^{'}=2c_s^2(F_{ty}u_y+F_{py}u_y)+c_s^2\hat{\Gamma}_{00}^p,\\
&&\hat{\sigma}_{11}^{'}=c_s^2(F_{tx}u_y+F_{ty}u_x)+c_s^2(F_{px}u_y+F_{py}u_x),\quad \hat{\sigma}_{21}^{'}=0,\quad \hat{\sigma}_{12}^{'}=0,\quad \hat{\sigma}_{22}^{'}=0.
\end{eqnarray*}
Using this, we then evaluate the various source moments projected to the orthogonal basis vectors and with a scaling based on the relaxation time for avoiding any spurious effects in the second order non-equilibrium moments as mentioned earlier, i.e., $\hat{m}^{s'}_j=\left(1-\frac{1}{2}\omega_j\right)\braket{K_j|S_\alpha}$, which yields
\begin{eqnarray*}
&&\hat{m}^{s'}_0=\left(1-\frac{1}{2}\omega_0\right)\hat{\sigma}_{00}^{'}, \quad \hat{m}^{s'}_1=\left(1-\frac{1}{2}\omega_1\right)\hat{\sigma}_{10}^{'},\quad \hat{m}^{s'}_2=\left(1-\frac{1}{2}\omega_2\right)\hat{\sigma}_{01}^{'},\\
&&\hat{m}^{s'}_3=\left(1-\frac{1}{2}\omega_3\right)\left[3(\hat{\sigma}_{20}^{'}+\hat{\sigma}_{02}^{'})-4\hat{\sigma}_{00}^{'}\right], \quad \hat{m}^{s'}_4=\left(1-\frac{1}{2}\omega_4\right)\left[\hat{\sigma}_{20}^{'}-\hat{\sigma}_{02}^{'}\right],\\
&&\hat{m}^{s'}_5=\left(1-\frac{1}{2}\omega_5\right)\hat{\sigma}_{11}^{'},\; \hat{m}^{s'}_6=\left(1-\frac{1}{2}\omega_6\right)\left[-3\hat{\sigma}_{21}^{'}+2\hat{\sigma}_{01}^{'}\right], \; \hat{m}^{s'}_7=\left(1-\frac{1}{2}\omega_7\right)\left[-3\hat{\sigma}_{12}^{'}+2\hat{\sigma}_{10}^{'}\right],\\
&&\hat{m}^{s'}_8=\left(1-\frac{1}{2}\omega_8\right)\left[9\hat{\sigma}_{22}^{'}-6(\hat{\sigma}_{20}^{'}+\hat{\sigma}_{02}^{'})-8\hat{\sigma}_{00}^{'}\right].
\end{eqnarray*}
Finally, by exploiting the orthogonal property of $\tensr{K}$ in $\delta g_\alpha^s = \tensr{K}^{-1}\hat{\mathbf{m}}^{s'}$, where $\hat{\mathbf{m}}^{s'}=\left(\tensr{I}-\frac{1}{2}\hat{\Lambda}\right)\tensr{K}\mathbf{S}$, with $\hat{\mathbf{m}}^{s'}=(\hat{m}^{s'}_0, \hat{m}^{s'}_1, \hat{m}^{s'}_2,\cdots,\hat{m}^{s'}_8)^\dag$, we get the modified cumulative source term due to various sources/forces in the cascaded LB scheme for two-phase flow as
\begin{eqnarray}
&&\delta g_0^s = \frac{1}{9}\left[\hat{m}_0^{s'}-\hat{m}_3^{s'}+\hat{m}_8^{s'}\right],\nonumber\\
&&\delta g_1^s = \frac{1}{36}\left[4\hat{m}_0^{s'}+6\hat{m}_1^{s'}-\hat{m}_3^{s'}+9\hat{m}_4^{s'}+6\hat{m}_7^{s'}-2\hat{m}_8^{s'}\right],\nonumber\\
&&\delta g_2^s = \frac{1}{36}\left[4\hat{m}_0^{s'}+6\hat{m}_2^{s'}-\hat{m}_3^{s'}-9\hat{m}_4^{s'}+6\hat{m}_6^{s'}-2\hat{m}_8^{s'}\right],\nonumber\\
&&\delta g_3^s = \frac{1}{36}\left[4\hat{m}_0^{s'}-6\hat{m}_1^{s'}-\hat{m}_3^{s'}+9\hat{m}_4^{s'}-6\hat{m}_7^{s'}-2\hat{m}_8^{s'}\right],\nonumber\\
&&\delta g_4^s = \frac{1}{36}\left[4\hat{m}_0^{s'}-6\hat{m}_2^{s'}-\hat{m}_3^{s'}-9\hat{m}_4^{s'}-6\hat{m}_6^{s'}-2\hat{m}_8^{s'}\right],\nonumber\\
&&\delta g_5^s = \frac{1}{36}\left[4\hat{m}_0^{s'}+6\hat{m}_1^{s'}+6\hat{m}_2^{s'}+2\hat{m}_3^{s'}+9\hat{m}_5^{s'}-3\hat{m}_6^{s'}-3\hat{m}_7^{s'}+\hat{m}_8^{s'}\right],\nonumber\\
&&\delta g_6^s = \frac{1}{36}\left[4\hat{m}_0^{s'}-6\hat{m}_1^{s'}+6\hat{m}_2^{s'}+2\hat{m}_3^{s'}-9\hat{m}_5^{s'}-3\hat{m}_6^{s'}+3\hat{m}_7^{s'}+\hat{m}_8^{s'}\right],\nonumber\\
&&\delta g_7^s = \frac{1}{36}\left[4\hat{m}_0^{s'}-6\hat{m}_1^{s'}-6\hat{m}_2^{s'}+2\hat{m}_3^{s'}+9\hat{m}_5^{s'}+3\hat{m}_6^{s'}+3\hat{m}_7^{s'}+\hat{m}_8^{s'}\right],\nonumber\\
&&\delta g_8^s = \frac{1}{36}\left[4\hat{m}_0^{s'}+6\hat{m}_1^{s'}-6\hat{m}_2^{s'}+2\hat{m}_3^{s'}-9\hat{m}_5^{s'}+3\hat{m}_6^{s'}-3\hat{m}_7^{s'}+\hat{m}_8^{s'}\right].\label{eq:cumulativesourcetermcomponents}
\end{eqnarray}

Next, the structure of the cascaded collision operator $(\tensr K\cdot \widehat {\mathbf h})_{\alpha}$ based on the discrete equilibrium central moments $\hat{\eta}_{mn}^{eq}$ given in Eq.~(\ref{eq:matchingmoments}) is determined as follows. For all non-conserved moments, i.e., for $(m+n)\geq 2$, we prescribe the relaxation of the discrete central moments $\hat{\overline{\eta}}_{mn}$ to their corresponding central moment equilibria $\hat{\eta}_{mn}^{eq}$ at a relaxation time $\omega_*$~\cite{geier2006cascaded,premnath2009incorporating}. That is, $\sum_\alpha (\tensr K\cdot \widehat {\mathbf h})_{\alpha} {{(e_{\alpha x}-u_x)}^m}{{(e_{\alpha y}-u_y)}^n}=\omega_*(\hat{\eta}_{mn}^{eq}-\hat{\overline{\eta}}_{mn})$. For the transformed distribution function $\overline{g}_\alpha$ employed in the cascaded LB scheme (Eq.~(\ref{eq:cascadedLBEtwophaseflowexplicit})), during a time step $\delta_t$, its zeroth moment change needs to be $\hat{\sigma}_{00}^{'}$, while its first order moments are required to change by $\hat{\sigma}_{10}^{'}$ and $\hat{\sigma}_{01}^{'}$ in order to consistently update the pressure field and the fluid momentum via the interfacial and body forces. On the other hand, the respective moment changes due to the sources given earlier are $\hat{m}^{s'}_0=\left(1-\frac{1}{2}\omega_0\right)\hat{\sigma}_{00}^{'}$, $\hat{m}^{s'}_1=\left(1-\frac{1}{2}\omega_1\right)\hat{\sigma}_{10}^{'}$, and $\hat{m}^{s'}_2=\left(1-\frac{1}{2}\omega_2\right)\hat{\sigma}_{01}^{'}$. Hence, to meet the above physical constraints, we effectively need to satisfy the following constraints: $\sum_\alpha (\tensr K\cdot \widehat {\mathbf h})_{\alpha}=\frac{\omega_0}{2}\hat{\sigma}_{00}^{'}$, $\sum_\alpha (\tensr K\cdot \widehat {\mathbf h})_{\alpha}e_{\alpha x}=\frac{\omega_1}{2}\hat{\sigma}_{10}^{'}$ and $\sum_\alpha (\tensr K\cdot \widehat {\mathbf h})_{\alpha}e_{\alpha y}=\frac{\omega_2}{2}\hat{\sigma}_{01}^{'}$. Based on these considerations for the lower order moment changes and the central moment relaxation for the higher order moments under collision mentioned above, the expressions for the components of the moment change vector $\mathbf{\widehat{h}}=(\widehat{h}_0,\widehat{h}_1,\widehat{h}_2,\ldots,\widehat{h}_8)^\dag$ can be determined, which are summarized as follows:
\begin{eqnarray}
&&\hat{h}_0=\frac{\omega_0}{2}\frac{\hat{\Gamma}_{00}^p}{9}, \quad \hat{h}_1=\frac{\omega_1}{2}\frac{c_s^2F_{tx}}{6}, \quad
\hat{h}_2=\frac{\omega_2}{2}\frac{c_s^2F_{ty}}{6},\nonumber\\
&&\hat{h}_3=\frac{\omega_3}{12}\left[2pc_s^2+\rho c_s^2(u_x^2+u_y^2)-(\hat{\overline{\eta}}_{20}^{'}+\hat{\overline{\eta}}_{02}^{'})\right],\nonumber\\
&&\hat{h}_4=\frac{\omega_4}{4}\left[\rho c_s^2(u_x^2-u_y^2)-(\hat{\overline{\eta}}_{20}^{'}-\hat{\overline{\eta}}_{02}^{'})\right],\nonumber\\
&&\hat{h}_5=\frac{\omega_5}{4}\left[\rho c_s^2u_xu_y-\hat{\overline{\eta}}_{11}^{'}\right],\nonumber\\
&&\hat{h}_6=\frac{\omega_6}{4}\left[\psi(\rho)(c_s^2+u_x^2)u_y+\hat{\overline{\eta}}_{21}^{'}-u_y\hat{\overline{\eta}}_{20}^{'}-2u_x\hat{\overline{\eta}}_{11}^{'}+3c_s^2\rho u_x^2u_y-u_x^2u_yp\right]\nonumber\\
&&\quad\quad\quad -u_y\left(\frac{3}{2}h_3+\frac{1}{2}\hat{h}_4\right)-2u_x\hat{h}_5,\nonumber\\
&&\hat{h}_7=\frac{\omega_7}{4}\left[\psi(\rho)(c_s^2+u_y^2)u_x+\hat{\overline{\eta}}_{12}^{'}-2u_y\hat{\overline{\eta}}_{11}^{'}-u_x\hat{\overline{\eta}}_{02}^{'}+3c_s^2\rho u_xu_y^2-u_xu_y^2p\right]\nonumber\\
&&\quad\quad\quad -u_x\left(\frac{3}{2}h_3-\frac{1}{2}\hat{h}_4\right)-2u_y\hat{h}_5,\nonumber\\
&&\hat{h}_8=\frac{\omega_8}{4}\left[c_s^6\rho+\psi(\rho)(c_s^2+u_x^2)(c_s^2+u_y^2)-\hat{\overline{\eta}}_{22}^{'}+2(u_y\hat{\overline{\eta}}_{21}^{'}+u_x\hat{\overline{\eta}}_{12}^{'})\right.\nonumber\\
&&\left.\quad\quad\quad -(u_y^2\hat{\overline{\eta}}_{20}^{'}+u_x^2\hat{\overline{\eta}}_{02}^{'})-4u_xu_y\hat{\overline{\eta}}_{11}^{'}+4c_s^2\rho u_x^2u_y^2-u_x^2u_y^2p\right]-2\hat{h}_3-\frac{1}{2}u_y^2(3\hat{h}_3+\hat{h}_4)\nonumber\\
&&\quad\quad\quad -\frac{1}{2}u_x^2(3\hat{h}_3-\hat{h}_4)-4u_xu_y\hat{h}_5-2u_y\hat{h}_6-2u_x\hat{h}_7.\label{eq:cascadedcollisionoperator}
\end{eqnarray}

Finally, the post-collision distribution function represented by $\widetilde{\overline{g}}_\beta$, where $\beta=0,1,2,\ldots, 8$, can be obtained by expanding $(\tensr K\cdot \widehat {\mathbf h})_{\alpha}$ in Eq.~(\ref{eq:cascadedLBEtwophaseflowexplicit}), which read as
\begin{eqnarray}
&&\widetilde{\overline{g}}_0 = \overline{g}_0+\left[\hat{h}_0-4(\hat{h}_3-\hat{h}_8)\right]+\delta g_0^s,\nonumber\\
&&\widetilde{\overline{g}}_1 = \overline{g}_1+\left[\hat{h}_0+\hat{h}_1-\hat{h}_3+\hat{h}_4+2(\hat{h}_7-\hat{h}_8)\right]+\delta g_1^s,\nonumber\\
&&\widetilde{\overline{g}}_2 = \overline{g}_2+\left[\hat{h}_0+\hat{h}_2-\hat{h}_3-\hat{h}_4+2(\hat{h}_6-\hat{h}_8)\right]+\delta g_2^s,\nonumber\\
&&\widetilde{\overline{g}}_3 = \overline{g}_3+\left[\hat{h}_0-\hat{h}_1-\hat{h}_3+\hat{h}_4-2(\hat{h}_7+\hat{h}_8)\right]+\delta g_3^s,\nonumber\\
&&\widetilde{\overline{g}}_4 = \overline{g}_4+\left[\hat{h}_0-\hat{h}_2-\hat{h}_3-\hat{h}_4-2(\hat{h}_6+\hat{h}_8)\right]+\delta g_4^s,\nonumber\\
&&\widetilde{\overline{g}}_5 = \overline{g}_5+\left[\hat{h}_0+\hat{h}_1+\hat{h}_2+2\hat{h}_3+\hat{h}_5-\hat{h}_6-\hat{h}_7+\hat{h}_8\right]+\delta g_5^s,\nonumber\\
&&\widetilde{\overline{g}}_6 = \overline{g}_6+\left[\hat{h}_0-\hat{h}_1+\hat{h}_2+2\hat{h}_3-\hat{h}_5-\hat{h}_6+\hat{h}_7+\hat{h}_8\right]+\delta g_6^s,\nonumber\\
&&\widetilde{\overline{g}}_7 = \overline{g}_7+\left[\hat{h}_0-\hat{h}_1-\hat{h}_2+2\hat{h}_3+\hat{h}_5+\hat{h}_6+\hat{h}_7+\hat{h}_8\right]+\delta g_7^s,\nonumber\\
&&\widetilde{\overline{g}}_8 = \overline{g}_8+\left[\hat{h}_0+\hat{h}_1-\hat{h}_2+2\hat{h}_3-\hat{h}_5+\hat{h}_6-\hat{h}_7+\hat{h}_8\right]+\delta g_8^s.\label{eq:postcollisioncascadedLBscheme}
\end{eqnarray}
This represents the collision step, and the streaming step then follows from rearranging Eq.~(\ref{eq:cascadedLBEtwophaseflowexplicit}) as
$\overline{g}_{\alpha}(\bm{x},t+\delta_t)=\widetilde{\overline{g}}_{\alpha}(\bm{x}-\bm{e}_\alpha\delta_t,t)$, where $\alpha=0,1,2,\cdots, 8$. Once the cascaded collision and streaming steps are performed, the two-phase flow fields, i.e., the hydrodynamic pressure and the velocity can be obtained via the zeroth and first moments of the transformed distribution function as
\begin{equation}
p=\sum_\alpha \overline{g}_\alpha+\frac{1}{2}\bm{F}_p\cdot\bm{u}\delta_t,\quad \rho c_s^2 \bm{u}=\sum_\alpha \overline{g}_\alpha\bm{e}_\alpha+\frac{1}{2}c_s^2\bm{F}_t\delta_t.\label{eq:hydrodynamicfields}
\end{equation}
Based on the Chapman-Enskog multiscale expansion (see e.g.,~\cite{premnath2009incorporating}), it can be shown that the above cascaded LB scheme represents the incompressible two-phase fluid motion, where the fluid's shear viscosity $\mu$ is related to the relaxation times of the second order moments as
\begin{equation}
\mu=\rho\nu=\rho c_s^2\left(\frac{1}{\omega_j}-\frac{1}{2}\right)\delta_t,\quad j=4,5, \label{eq:fluidviscosity}
\end{equation}
and the rest of the relaxation times, which can influence numerical stability, are set to unity in this work. It may be noted that in the implementation of our cascaded LB formulation, all the spatial gradients of the phase field variable $\phi$ required in the computation of the interfacial normal $\bm{n}=(n_x,n_y)$ and the surface tension force $\bm{F}_s$ are obtained using a second order isotropic finite difference scheme. In addition, in view of Eq.~(\ref{eq:sfluidpropertiesinterpolation}), the spatial gradients of the density $\rho$ are directly expressed in terms of those of $\phi$. The solution procedure for the evolution of the phase field will be discussed next.

\section{Cascaded LB Method for Solution of Phase-Field based Interfacial Dynamics\label{sec:cascadedLBMinterfacialmotion}}
We will now construct another cascaded LB scheme for the solution of the conservative Allen-Cahn equation (ACE) given in Eq.~(\ref{eq:ACE}). Since the ACE is a convection-diffusion equation (CDE) with an additional interface sharpening flux term, our solution approach is based on modifying the central moment cascaded approach that we developed recently for CDE in a MRT formulation~\cite{hajabdollahi2018sym,hajabdollahi2018central,hajabdollahi2019cascaded}, where this additional term is included in the first order moment equilibria. This strategy is an extension of the approach proposed in~\cite{geier2015conservative}. In this regard, we consider a D2Q9 lattice using the same orthogonal moment basis vectors and the matrix given in Eqs.~(\ref{eq:orthogonalbasisvectors}) and (\ref{eq:collisionmatrix}), respectively.

Then, the collision and streaming steps of such a cascaded LB scheme for the evolution of the discrete distribution function $f_{\alpha}$ can be respectively represented as
\begin{subequations}
\begin{eqnarray}
\widetilde{f}_{\alpha}(\bm{x},t)&=&f_{\alpha}(\bm{x},t)+(\tensr K\cdot \widehat {\mathbf g})_{\alpha,}\label{eq:ACEcollisionLB}\\
f_\alpha(\bm{x},t+\delta_t)&=&\widetilde{f}_\alpha(\bm{x}-\bm{e}_\alpha \delta_t,t).
\label{eq:ACEstreamingLB}
\end{eqnarray}
\end{subequations}
In order to design a cascaded collision operator to solve for the phase field variable $\phi$ described by an conservative Allen-Cahn equation (ACE), we first define the following central moments and raw moments of the distribution function $f_\alpha$ and its equilibrium $f_\alpha^{eq}$, respectively, as
\begin{eqnarray}
\left( {\begin{array}{*{20}{l}}
{{{\hat \kappa }_{mn}}}\\
{\hat {\kappa} _{mn}^{eq}}
\end{array}} \right) = \sum\limits_\alpha  {\left( {\begin{array}{*{20}{l}}
{{f_\alpha}}\\
{f_\alpha ^{eq}}
\end{array}} \right)} {{(e_{\alpha x}-u_x)}^m}{{(e_{\alpha y}-u_y)}^n},
\label{eq:discretecentralmomentsACE}
\end{eqnarray}
\begin{eqnarray}
\left( {\begin{array}{*{20}{l}}
{{{\hat \kappa }_{mn}^{'}}}\\
{\hat {\kappa} _{mn}^{eq '}}
\end{array}} \right) = \sum\limits_\alpha  {\left( {\begin{array}{*{20}{l}}
{{f_\alpha}}\\
{f_\alpha ^{eq}}
\end{array}} \right)} {{e_{\alpha x}^m}}{{e_{\alpha y}^n}}.\label{eq:discreterawmomentsACE}
\end{eqnarray}
Then, we consider the continuous central moments of the equilibria
\begin{equation}
\widehat{\Pi}^{{eq,\phi }}_{mn}=\int_{-\infty}^{\infty}\int_{-\infty}^{\infty}f^{eq}(\xi_x-u_x)^m(\xi_y-u_y)^n d\xi_xd\xi_y\label{eq:continuouscentralmomentsequilibriaACE}
\end{equation}
by defining the equilibrium distribution function $f^{{eq}}$ in analogy with the local Maxwell distribution function by replacing the density with the phase field variable $\phi$: $f^{eq}\equiv f^{eq}(\phi,\bm{u},\bm{\xi})=\frac{\phi}{2\pi c_{s\phi}^2}\exp\left[-\frac{\left(\bm{\xi}-\bm{u}\right)^2}{2c_{s\phi}^2}\right]$. Here $c_{s\phi}$ is a free parameter, which will be related to the coefficient of diffusivity $M_\phi$ later. Typically, we set $c^2_{s\phi}=\frac{1}{3}$. The relaxation of the central moments to the corresponding equilibria given above only models a diffusion process. In order to account for the counteracting phase separation flux components $-\theta n_x$ and $-\theta n_y$ appearing in the conservative ACE (Eq.~(\ref{eq:ACE})), where $\bm{n}=(n_x,n_y)$ is the interface normal, we modify the first order continuous central moments from being null to $\widehat{\Pi}^{{eq,\phi }}_{10}=M_{\phi}\theta n_x$ and $\widehat{\Pi}^{{eq,\phi }}_{01}=M_{\phi}\theta n_y$. Then, by matching of the discrete and continuous central moments of the equilibria, i.e., $\widehat{\kappa}^{eq}_{mn}=\widehat{\Pi}^{{eq,\phi }}_{mn}$ for all the nine independent moments supported by the lattice, we obtain the components of $\widehat{\kappa}^{eq}_{mn}$ as
\begin{eqnarray*}
&&\hat{\kappa}_{00}^{eq}=\phi,\quad \hat{\kappa}_{10}^{eq}=M_\phi\theta n_x,\quad \hat{\kappa}_{01}^{eq}=M_\phi\theta n_y,\quad \hat{\kappa}_{20}^{eq}=c_{s\phi}^2\phi,\quad \hat{\kappa}_{02}^{eq}=c_{s\phi}^2\phi,\quad \hat{\kappa}_{11}^{eq}=0,\\
&&\hat{\kappa}_{21}^{eq}=0,\quad \hat{\kappa}_{12}^{eq}=0,\quad \hat{\kappa}_{22}^{eq}=c_{s\phi}^4\phi.
\end{eqnarray*}
The cascaded collision operator can then be constructed by prescribing the relaxation of central moments of different orders to their equilibria, i.e., $\sum_\alpha (\tensr K\cdot \widehat {\mathbf g})_{\alpha} {{(e_{\alpha x}-u_x)}^m}{{(e_{\alpha y}-u_y)}^n}=\omega_*^\phi(\hat{\kappa}_{mn}^{eq}-\hat{\kappa}_{mn})$, where only the zeroth moment being conserved ($\hat{\kappa}_{00}=\hat{\kappa}_{00}^{eq}=\phi$), and $\omega_*^\phi$ are the various relaxation times. The resulting changes in all the nine components of moments under collision, i.e., $\mathbf{\hat{g}}=(\hat{g}_0,\hat{g}_1,\hat{g}_2,\cdots,\hat{g}_8)$ can be summarized as follows:
\begin{eqnarray}
&&\hat{g}_0=0, \quad \hat{g}_1=\frac{\omega_1^\phi}{6}\left[\phi u_x+\boxed{M_\phi\theta n_x}-\hat{\kappa}_{10}^{'}\right],\quad \hat{g}_2=\frac{\omega_2^\phi}{6}\left[\phi u_y+\boxed{M_\phi\theta n_y}-\hat{\kappa}_{01}^{'}\right],\nonumber\\
&&\hat{g}_3=\frac{\omega_3^\phi}{12}\left[2c_{s\phi}^2\phi-(u_x^2+u_y^2)\phi-(\hat{\kappa}_{20}^{'}+\hat{\kappa}_{02}^{'})+2(u_x\kappa_{10}^{'}+u_y\hat{\kappa}_{01}^{'})\right]+u_x\hat{g}_1+u_y\hat{g}_2,\nonumber\\
&&\hat{g}_4=\frac{\omega_4^\phi}{4}\left[-(u_x^2-u_y^2)\phi-(\hat{\kappa}_{20}^{'}-\hat{\kappa}_{02}^{'})+2(u_x\kappa_{10}^{'}-u_y\hat{\kappa}_{01}^{'})\right]+3(u_x\hat{g}_1-u_y\hat{g}_2),\nonumber\\
&&\hat{g}_5=\frac{\omega_5^\phi}{4}\left[-u_xu_y\phi-\hat{\kappa}_{11}^{'}+(u_x\kappa_{01}^{'}+u_y\hat{\kappa}_{10}^{'})\right]+\frac{3}{2}(u_x\hat{g}_2+u_y\hat{g}_1),\nonumber\\
&&\hat{g}_6=\frac{\omega_6^\phi}{4}\left[-u_x^2u_y\phi+\hat{\kappa}_{21}^{'}-u_y\kappa_{20}^{'}-2u_x\kappa_{11}^{'}+2u_xu_y\kappa_{10}^{'}+u_x^2\kappa_{01}^{'}\right]+3u_xu_y\hat{g}_1\nonumber\\
&&\quad\quad\quad +\left(\frac{3}{2}u_x^2+1\right)\hat{g}_2-\frac{3}{2}u_y\hat{g}_3-\frac{1}{2}u_y\hat{g}_4-2u_x\hat{g}_5,\nonumber\\
&&\hat{g}_7=\frac{\omega_7^\phi}{4}\left[-u_xu_y^2\phi+\hat{\kappa}_{12}^{'}-u_x\kappa_{02}^{'}-2u_y\kappa_{11}^{'}+2u_xu_y\kappa_{01}^{'}+u_y^2\kappa_{10}^{'}\right]+\left(\frac{3}{2}u_y^2+1\right)\hat{g}_1\nonumber\\
&&\quad\quad\quad +3u_xu_y\hat{g}_2-\frac{3}{2}u_y\hat{g}_3+\frac{1}{2}u_x\hat{g}_4-2u_y\hat{g}_5,\nonumber\\
&&\hat{g}_8=\frac{\omega_8^\phi}{4}\left[c_{s\phi}^4\phi-\hat{\kappa}_{22}^{'}+2(u_x\hat{\kappa}_{12}^{'}+u_y\hat{\kappa}_{21}^{'})-(u_y^2\hat{\kappa}_{20}^{'}+u_x^2\hat{\kappa}_{02}^{'})-4u_xu_y\hat{\kappa}_{11}^{'}+2(u_xu_y^2\hat{\kappa}_{10}^{'}+u_x^2u_y\hat{\kappa}_{01}^{'})\right.\nonumber\\
&&\quad\quad\quad\left.-u_x^2u_y^2\phi\right] + (2u_x+3u_xu_y^2)\hat{g}_1+(2u_y+3u_x^2u_y)\hat{g}_2-(2+\frac{3}{2}(u_x^2+u_y^2))\hat{g}_3+\frac{1}{2}(u_x^2-u_y^2)\hat{g}_4\nonumber\\
&&\quad\quad\quad -4u_xu_y\hat{g}_5-2u_y\hat{g}_6-2u_x\hat{g}_7,
\end{eqnarray}
where the relaxation times of the first order moments $\omega_1^{\phi}$ and $\omega_2^{\phi}$ are related to the mobility parameter $M_\phi$ in Eq.~(\ref{eq:ACE}) via $M_\phi=c_{s\phi}^2\left(\frac{1}{\omega_j^\phi}-\frac{1}{2}\right)\delta_t, j=1,2$, and the rest of the relaxation times are set to unity.
Finally, the post-collision distribution function $\widetilde{f}_{\alpha}$ can be explicitly written after expanding $(\tensr K\cdot \widehat {\mathbf g})_{\alpha}$ in Eq.~(\ref{eq:ACEcollisionLB}) as
\begin{eqnarray}
&&\widetilde{f}_0 = f_0+\left[\hat{g}_0-4(\hat{g}_3-\hat{g}_8)\right],\nonumber\\
&&\widetilde{f}_1 = f_1+\left[\hat{g}_0+\hat{g}_1-\hat{g}_3+\hat{g}_4+2(\hat{g}_7-\hat{g}_8)\right],\nonumber\\
&&\widetilde{f}_2 = f_2+\left[\hat{g}_0+\hat{g}_2-\hat{g}_3-\hat{g}_4+2(\hat{g}_6-\hat{g}_8)\right],\nonumber\\
&&\widetilde{f}_3 = f_3+\left[\hat{g}_0-\hat{g}_1-\hat{g}_3+\hat{g}_4-2(\hat{g}_7+\hat{g}_8)\right],\nonumber\\
&&\widetilde{f}_4 = f_4+\left[\hat{g}_0-\hat{g}_2-\hat{g}_3-\hat{g}_4-2(\hat{g}_6+\hat{g}_8)\right],\nonumber\\
&&\widetilde{f}_5 = f_5+\left[\hat{g}_0+\hat{g}_1+\hat{g}_2+2\hat{g}_3+\hat{g}_5-\hat{g}_6-\hat{g}_7+\hat{g}_8\right],\nonumber\\
&&\widetilde{f}_6 = f_6+\left[\hat{g}_0-\hat{g}_1+\hat{g}_2+2\hat{g}_3-\hat{g}_5-\hat{g}_6+\hat{g}_7+\hat{g}_8\right],\nonumber\\
&&\widetilde{f}_7 = f_7+\left[\hat{g}_0-\hat{g}_1-\hat{g}_2+2\hat{g}_3+\hat{g}_5+\hat{g}_6+\hat{g}_7+\hat{g}_8\right],\nonumber\\
&&\widetilde{f}_8 = f_8+\left[\hat{g}_0+\hat{g}_1-\hat{g}_2+2\hat{g}_3-\hat{g}_5+\hat{g}_6-\hat{g}_7+\hat{g}_8\right].
\end{eqnarray}
This is followed by performing the streaming step shown in Eq.~(\ref{eq:ACEstreamingLB}), which then updates the phase field variable $\phi$
via taking the zeroth moment of $f_{\alpha}$ as
\begin{equation}
\phi=\sum_\alpha f_{\alpha}.
\end{equation}

\section{\label{sec:resultsdiscussion} Results and Discussion}
We will now present a validation study of our new cascaded LB approach developed for incompressible two-phase flows for a variety of benchmark
problems with surface tension effects. Since the LB formulation for the interface capturing based on the conservative ACE has been analyzed in Ref.~\cite{geier2015conservative}, we will limit the validation of our implementation in this regard for one benchmark problem below (Sec.~\ref{sec:resultcircularinterfaceshearflow}). Instead, most of our focus in what follows will be on investigating the cascaded LB methods presented in the previous two sections for the coupled solution of the two-phase fluid motion with interfacial dynamics, especially at high density ratios and under different interfacial flow configurations.

\subsection{\label{sec:resultcircularinterfaceshearflow} Evolution of a circular interface in imposed shear flow}
\begin{figure}[htbp]
\centering
    \subfloat[\label{subfig-2:dummy}]{
        \includegraphics[scale=.75]{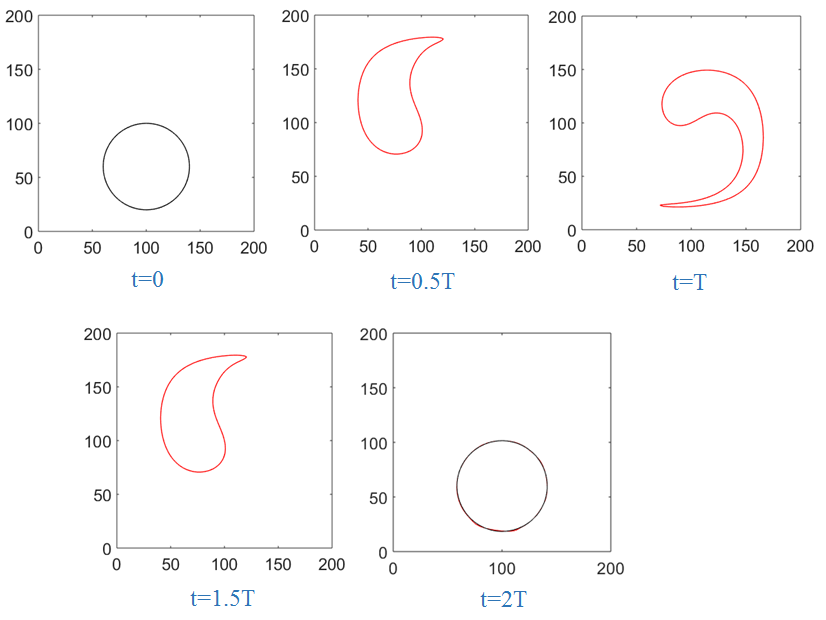}
         }
    \caption{Snapshots of the  interface under an imposed shear flow with an initially circular shape computed by the cascaded LB method.}
    \label{fig:circular}
\end{figure}

We will first assess the ability of the cascaded LB scheme for the solution of the conservative ACE (see Sec.~\ref{sec:cascadedLBMinterfacialmotion}) to capture the kinematical effects of the interfacial motion under deformation and rotational effects with good fidelity. In this regard, we consider a circular interface subjected to an imposed shear flow given by the following velocity field in a periodic square domain of size $L_0$~\cite{rudman1997volume}
\begin{eqnarray*}
u_x(x,y)&=&-U_0\pi \cos{\left[\pi(x/L_0-1/2)\right]}\sin{\left[\pi(y/L_0-1/2)\right]}\\
u_y(x,y)&=&U_0\pi \sin{\left[\pi(x/L_0-1/2)\right]}\cos{\left[\pi(y/L_0-1/2)\right]},
\end{eqnarray*}
where $U_0$ is the velocity scale. In our simulations, we take the radius of the circular interface to be $R = L_0/5$, whose center is
initially located at $(x_c,y_c) = (L_0/2,3L_0/10)$ in a square computational domain resolved with  $L_0 = 200$. Moreover, the numerical
parameters of the conservative ACE, i.e., the width $W$ and the mobility $M_\phi$ are set as follows: $W=3$ and latter is obtained by
considering a Peclet number $Pe=U_0W/M_\phi=60$. To guide interface undergoing deformation and rotation to return to its original position at $T = 2T_f$, where $T_f=L_0/U_0$, the velocity field given above is reversed at $T = T_f$. Figure~\ref{fig:circular} presents snapshots of the interface, identified by the contours of $(\phi_A+\phi_B)/2$ at the instants $T = 0, 0.5T_f, T_f, 1.5T_f, 2T_f$. It can be seen that the interface undergoes advection with complex shape changes under shear, and the cascaded LB method faithfully recovers the original circular shape with good accuracy after completing
a cycle.

\subsection{\label{sec:resultLaplaceYoungrelation} Laplace-Young relation of a static drop}
\begin{figure}[htbp]
\centering
    \subfloat[\label{subfig-2:dummy}]{
        \includegraphics[height=7.5cm, width=15cm]{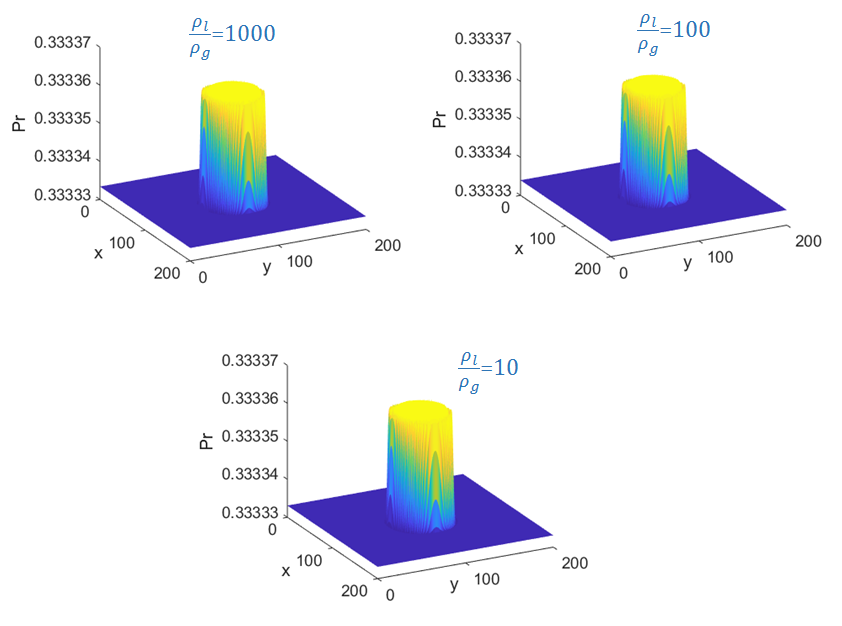}
         }
                                     \caption{Surface contours of the pressure distribution of a single static drop of radius $R=30$ at different density ratios $\rho_A/\rho_B$  with surface tension $\sigma=1\times 10^{-3}$ in a periodic square domain.}
    \label{fig:laplace}
\end{figure}

\begin{figure}[htbp]
\centering
    \subfloat[\label{subfig-2:dummy}]{
        \includegraphics[height=10cm, width=10cm]{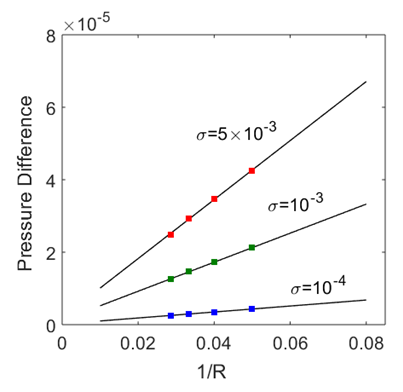}
         }
                                     \caption{Comparison of the computed pressure differences (symbols) obtained using the cascaded LB method against the analytical predictions using the Laplace-Young relation for various values of the drop curvature $1/R$ with surface tension $\sigma=5\times 10^{-3},1\times 10^{-3}, 1\times 10^{-4}$.}
    \label{fig:laplace1}
\end{figure}

We will now make a quantitative verification of the ability of the coupled cascaded LB formulations in the computation of the various forces and their balances in a static drop immersed in a fluid medium by considering high density ratios. In this regard, according to the analytical predictions
of the Laplace-Young's relation, for a 2D drop at rest, the pressure difference between the drop and the ambient fluid ($\Delta P$) is related to the
surface tension $\sigma$ and its radius of curvature ($1/R$) via $\Delta P = \sigma/R$, which we will use for comparison. In the simulations, we consider a drop of density $\rho_A$ surrounded by an ambient fluid of density $\rho_B$ and placed in the center of a periodic square domain resolved by $200\times 200$ grid nodes. We first performed simulations with a drop of radius $R=30$ by considering a surface tension $\sigma=1\times 10^{-3}$ at various density ratios of $\rho_A/\rho_B=10, 100, 1000$ till they reached equilibrium. Figure~\ref{fig:laplace} shows the surface contours of
the pressure differences between the drop and the ambient fluid. It is evident that the pressure distribution within the drop is smooth and uniform
and the jump across the interface is sharp and independent of the density ratio as expected. The cascaded LB method is seen to be robust even at relatively high density contrasts.
Then, Fig.~\ref{fig:laplace1} shows a comparison between the computed pressure differences between the drop and the ambient fluid as a function of its curvature for three different values of the surface tension $\sigma = 1\times 10^{-4}, 1\times 10^{-3}$, and $5\times 10^{-3}$ at a density ratio of $1000$ against the predictions given by the Laplace-Young relation. It verifies the expected linear dependence between $\Delta P$ and $1/R$ and
the computed results are found to be in good quantitative agreement with the analytical solution.
\subsection{\label{sec:resultRTinstability}Rayleigh-Taylor instability}
Next, we will investigate the cascaded LB methods for simulation of the classical Rayleigh-Taylor (R-T) instability. Such a gravitational acceleration-driven instability arises when a heavier fluid of density $\rho_A$ is placed on top of a lighter fluid of density $\rho_B$ in the presence of gravity, and the interface between the two fluids undergoes complex unsteady motion. A mesh size of $L\times 4L$,  where $L=201$, is employed, and periodic boundary conditions along the lateral vertical sides and no-slip boundary conditions at the top and bottom boundaries are imposed. The initial perturbation at the interface between the two fluids to initiate instability is described by a cosinusoidal function given by $ y_0=2L+0.1L \cos(2\pi x/L)$, where the origin of the coordinate system is fixed at the left bottom corner of the computational domain. The interfacial instability is characterized by the Reynolds number $\mbox{Re}=\rho_A \sqrt{gL}L/\mu$ based on a velocity scale $U_c=\sqrt{gL}$, and the Atwood number $\mbox{At}=(\rho_A-\rho_B)/(\rho_A+\rho_B)$. Here, $\mu$ is the dynamic viscosity and $g$ is the acceleration due to gravity. The dimensionless timescale $T$ is then defined based on $U_c$ and $L$ as $T=U_c/(L\sqrt{\mbox{At}})$. In addition, for interface capturing, we consider $W=5$, and the Peclet number $\mbox{Pe}=U_cL/M_\phi=3000$.

\begin{figure}[htbp]
\centering
    \subfloat[\label{subfig-2:dummy}]{
        \includegraphics[height=6.8cm, width=15cm]{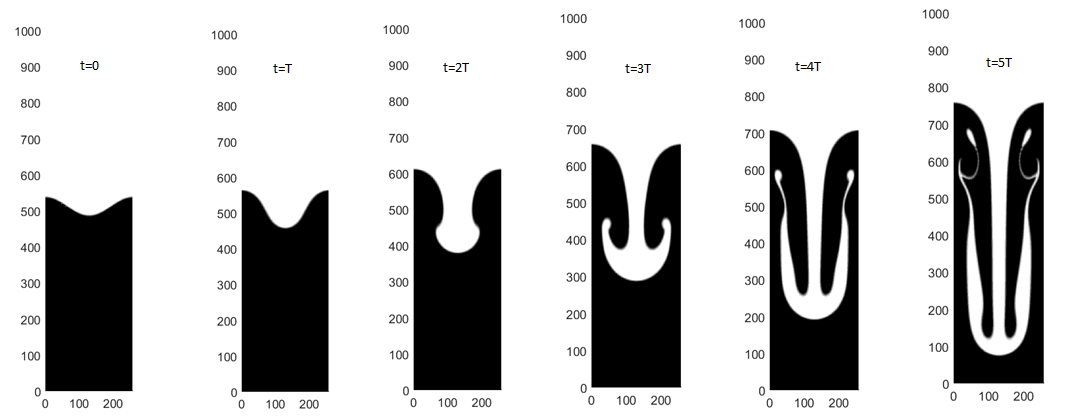}
         }
                               \\
    \subfloat[\label{subfig-2:dummy}]{
        \includegraphics[height=6.8cm, width=15cm] {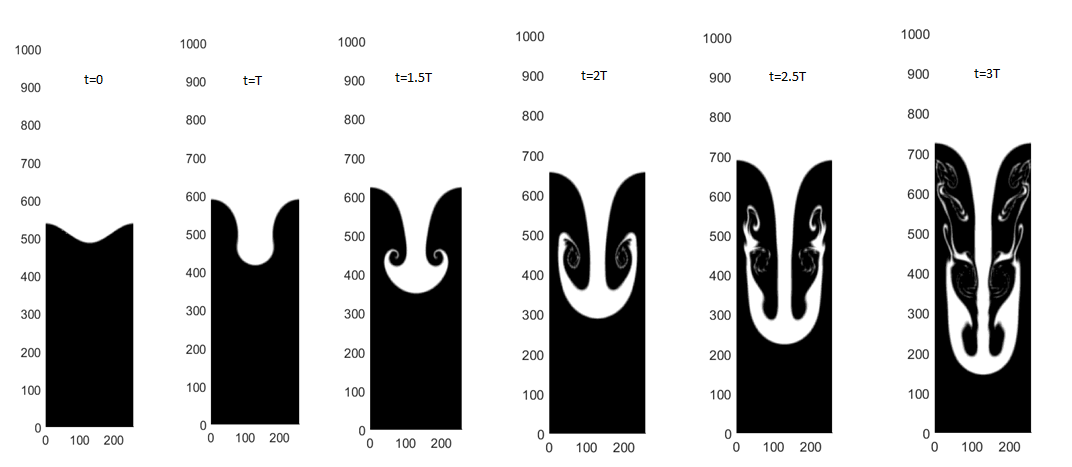}
         }
         \caption{Snapshots of simulation of Rayleigh-Taylor instability at $\mbox{At} = 0.5$ and (a) $\mbox{Re} = 256$ and (b) $\mbox{Re} = 3000$.}
         \label{fig:rising}
\end{figure}

\begin{figure}[htbp]
\centering
    \subfloat[\label{subfig-2:dummy}]{
        \includegraphics[height=6.8cm, width=7cm]{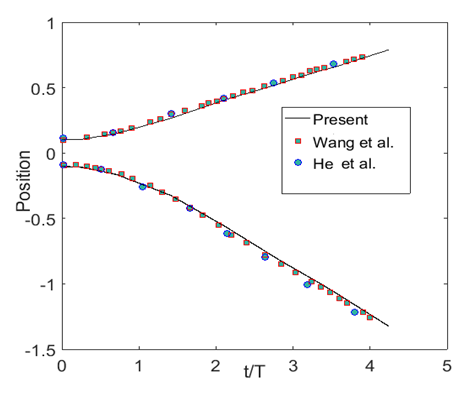}
         }
    \subfloat[\label{subfig-2:dummy}]{
        \includegraphics[height=6.8cm, width=7cm] {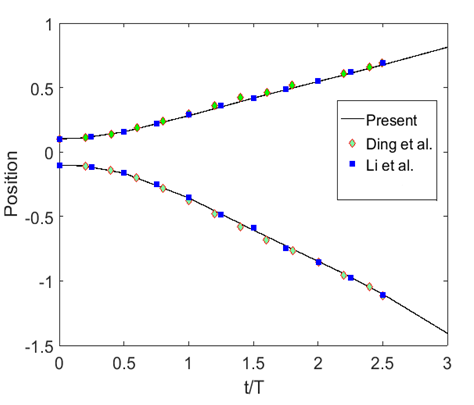}
         }
    \caption{Time evolution of the positions of  the bubble front and  the spike tip for Rayleigh-Taylor instability at $\mbox{At}=0.5$ and (a) $\mbox{Re}=256$ and (b) $\mbox{Re}=3000$.}
    \label{fig:rising1}
\end{figure}

By fixing $\mbox{At}=0.5$, we performed simulations for two cases of the Reynolds number, i.e., $\mbox{Re}=256$ and $3000$. Figure~\ref{fig:rising} presents the evolution of the interface under flow instability at these two Reynolds numbers. In general, the spike formation by the heavier fluid
moving downward is accompanied by a bubble of the lighter fluid rising upwards. The interface between the fluids undergoes complex shape changes leading to a roll-up of its tails under the dynamical effects of the two moving fluids. Moreover, at higher $\mbox{Re}$, when the inertial effects
predominate over the viscous effects, small scale flow structures emerge. The snapshots of the simulated results of the R-T instability at various time instants are in overall agreement with the prior numerical results at $\mbox{Re}=256$ (e.g.,~\cite{he1999lattice,wang2015mass}) and $\mbox{Re}=3000$ (e.g.,~\cite{ding2007diffuse,li2012additional}).
Moreover, Fig.~\ref{fig:rising1} shows quantitative comparisons of the computed values of the non-dimensional locations of the spike and bubble fronts scaled by $L$ at both $\mbox{Re}$ against prior numerical reference data. It can be that the numerical results obtained using the cascaded LB formulations for time evolution of the interface locations evaluated at the center (spike) and at the edges (bubble) are in good quantitative agreement with the respective reference results at both $\mbox{Re}=256$ and $\mbox{Re}=3000$.

\subsection{\label{sec:resultfallingdrop} Falling drop under gravity}
We will now consider another unsteady two-phase flow problem involving a drop falling under a gravitational field. In such a case, during the
descent of the drop, it undergoes significant shape changes due to deformation, which arises from a complex interplay between the gravity for
force, surface tension force and the viscous force. A drop of diameter $D=30$ with a density $\rho_A$ is placed initially at a location of $(75,300)$ in a rectangular domain that is divided into $151 \times 451$ lattice nodes (with the origin of the coordinate system being located at the left bottom corner), and filled with a lighter ambient fluid of density $\rho_B$. Free-slip boundary conditions are imposed on the top and bottom boundaries and lateral vertical sides are taken to be periodic. For this computational set up, the gravitational force is applied everywhere by setting $\bm{F}_{ext}=-(\rho-\rho_B)g \bm{j}$. The drop dynamics is characterized by the following non-dimensional numbers: Eotvos number $\mbox{Eo}=g(\rho_A-\rho_B)D^2/\sigma$ representing the gravity force relative to the surface tension and the Ohnesorge number $\mbox{Oh}=\mu_A/\sqrt{\rho_A D \sigma}$ representing the viscous effects. Following Ref.~\cite{fakhari2009simulation}, we fix $\rho_A/\rho_B=5$, $\mbox{Eo}=43$ and study the influence of $\mbox{Oh}$ by considering $\mbox{Oh}=0.3, 0.7$ and $1.0$, with $\nu_A=\nu_B=\nu$. These three values
of $\mbox{Oh}$ are obtained by setting $\nu=0.1, 0.2333$ and $0.3333$, respectively. For reporting results, the instantaneous time $t$ is non-dimensionalized as $T=t/\sqrt{D/g}$.

\begin{figure}[htbp]
\centering
    \subfloat[\label{subfig-2:dummy}]{
        \includegraphics[height=7.5cm, width=15cm]{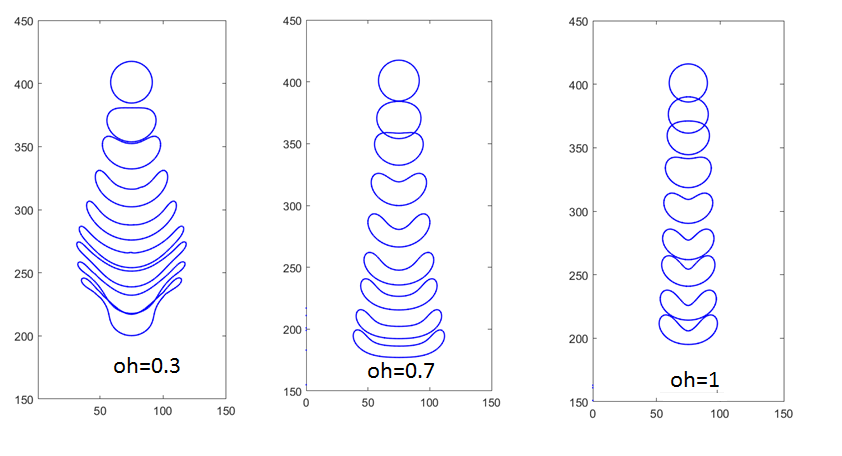}
         }
         \caption{Evolution of a deforming drop falling under gravity for various values of the Ohnesorge number $\mbox{Oh}$ of 0.3, 0.7 and 1.0 at a fixed Eotvos number $\mbox{Eo}=43$ shown at time instants $T = 0, 2.04,3.05, 4.07, 5.09, 6.11, 7.13, 8.14$, and $9.16$ (from top to bottom).}
    \label{fig:Falling}
\end{figure}

Figure~\ref{fig:Falling} presents the snapshots of the evolution of the interface of the falling drop for the above three cases of $\mbox{Oh}$. In general, it can be seen that as $Oh$ increases, the viscous force increases relative to the surface tension force and hence the drop deformation
is reduced. Thus, at a large value of $\mbox{Oh}=1.0$, the drop undergoes relative small deformation attaining a steady state, while at $\mbox{Oh}=0.7$, it is stretched more along the horizontal direction by the surface tension force after initially taking an ellipsoidal shape. On the other hand, at a still lower $\mbox{Oh}=0.3$, the drop becomes considerably slender along the sides, while exhibiting bag-like shape due to shear under gravity in the presence of the prevailing surface tension force with smaller viscous force effects at later stages. These computed drop shape variations at different times with $\mbox{Oh}$ are consistent with the findings reported in Ref.~\cite{fakhari2009simulation}.

\subsection{\label{sec:resultrisingbubble}Buoyancy-driven rising bubble}
Next, we examine the ability of our cascaded LB formulations to simulate a well-defined two-phase flow problem involving a moving dispersed phase in a continuous phase with high density contrasts than those considered in the previous two cases. In this regard, we consider a bubble of diameter $D$
and density $\rho_B$ rising in an ambient fluid of density $\rho_A$, with $\rho_A/\rho_B$ being 1000, by buoyancy forces under various parametric conditions. This represents the buoyant motion of an air bubble in water and is of practical interest. Our goal is to test the robustness of the cascaded LB approach to capture the various shape changes the bubble undergoes due to the balance between the different competing forces as well as simulate the time history of the bubble path with quantitative accuracy.

The computational configuration consists of a rectangular domain with a grid resolution of $161 \times 481$ in which a bubble of diameter resolved with $64$ grid nodes is initially centered at a location $(40,120)$ (with the coordinate system's origin being situated at the bottom left corner of the domain). Free slip boundary conditions are imposed on the two vertical sides and the no-slip conditions are considered on the top and bottom boundaries. This set up corresponds to that discussed in Refs.~\cite{hysing2009quantitative,aland2012benchmark}. The bubble is set in motion by applying a body force given by $\bm{F}_{ext}=-(\rho-\rho_A)g\bm{j}$. The characteristic scales of this two-phase flow problem are: the length scale $L=D$, the velocity scale $U_g=\sqrt{gD}$, which represents the gravitational velocity, and the time scale $T=L/U_g$. Based on these and the various competing forces (i.e., buoyancy, viscous and surface tension), the non-dimensional parameters of this two-phase flow problem are the Reynolds number $\mbox{Re}=\rho_AU_gD/\mu_A$ and the Eotvos number $\mbox{Eo}=\rho_AU_g^2D/\sigma$, along with the ratios of the fluid properties $\rho_A/\rho_B$ and $\mu_A/\mu_B$. The non-dimensional time for reporting time histories is represented by $t^*=t/T$. Depending on the magnitudes of these dimensionless groups, the bubble undergoes complex interfacial shape changes, attaining either spherical-cap, dimpled ellipsoidal-cap or skirted configurations, among various possibilities~\cite{clift1978bubbles}.

\begin{figure}[htbp]
\centering
    \subfloat[\label{subfig-2:dummy} Eo=10]{
        \includegraphics[height=6.5cm, width=15cm]{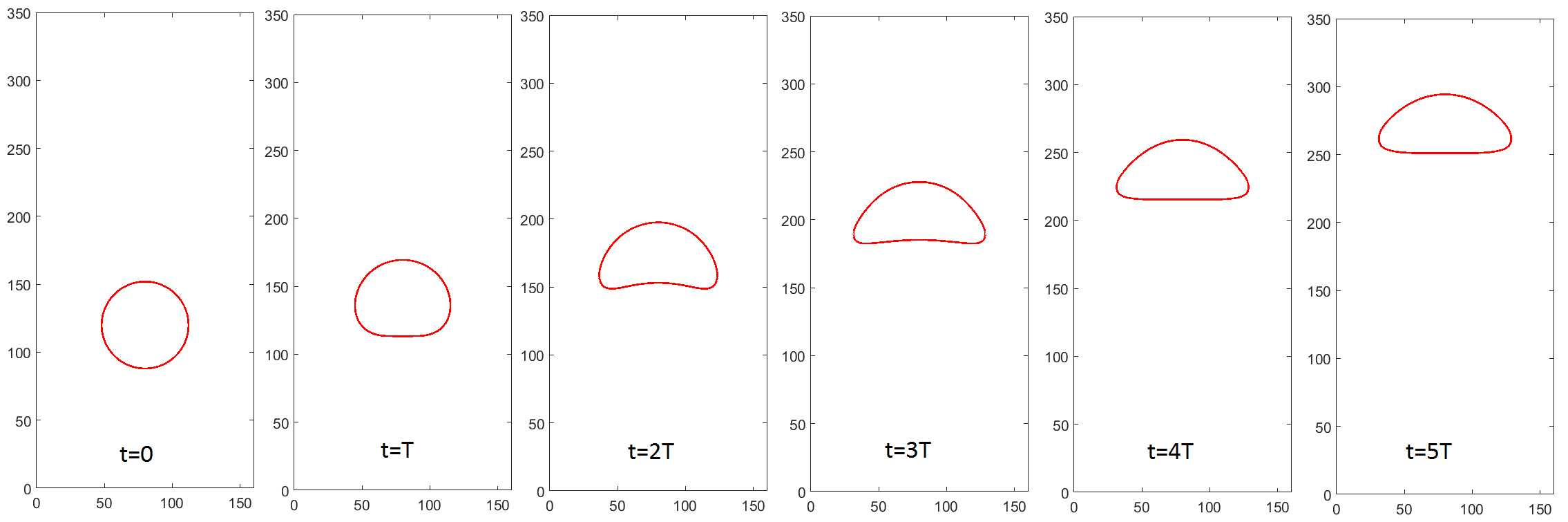}
         }
                               \\
    \subfloat[\label{subfig-2:dummy} Eo=50]{
        \includegraphics[height=6.5cm, width=15cm] {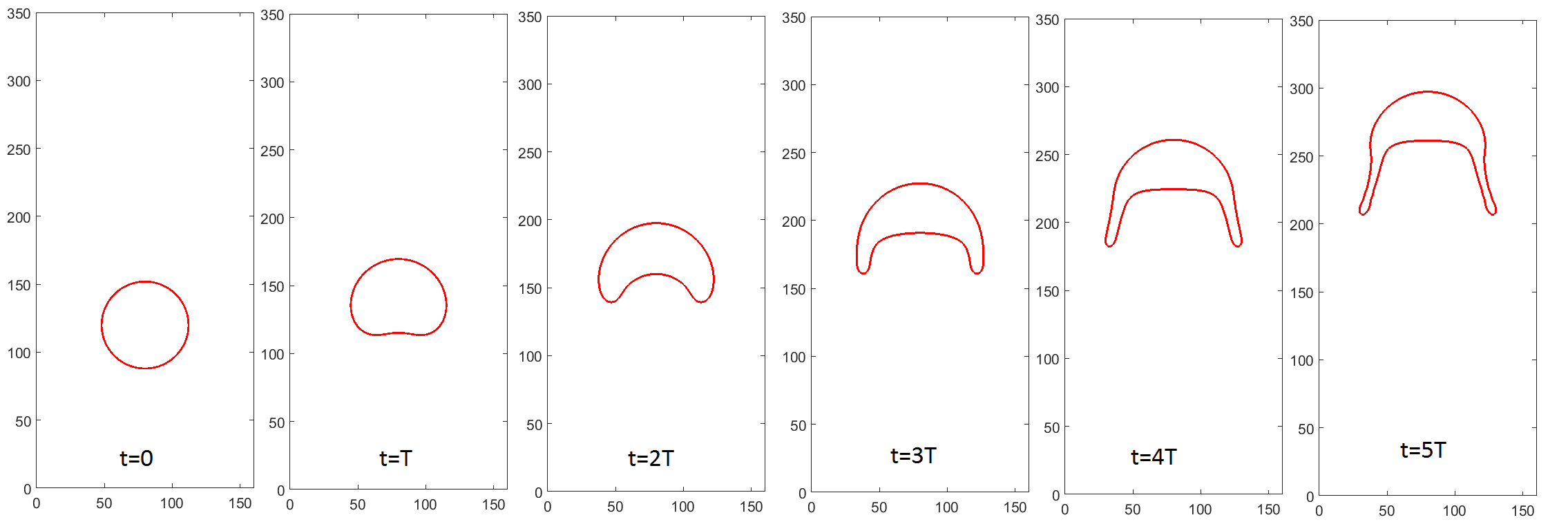}
         }
               \\
    \subfloat[\label{subfig-2:dummy} Eo=125]{
        \includegraphics [height=6.5cm, width=15cm] {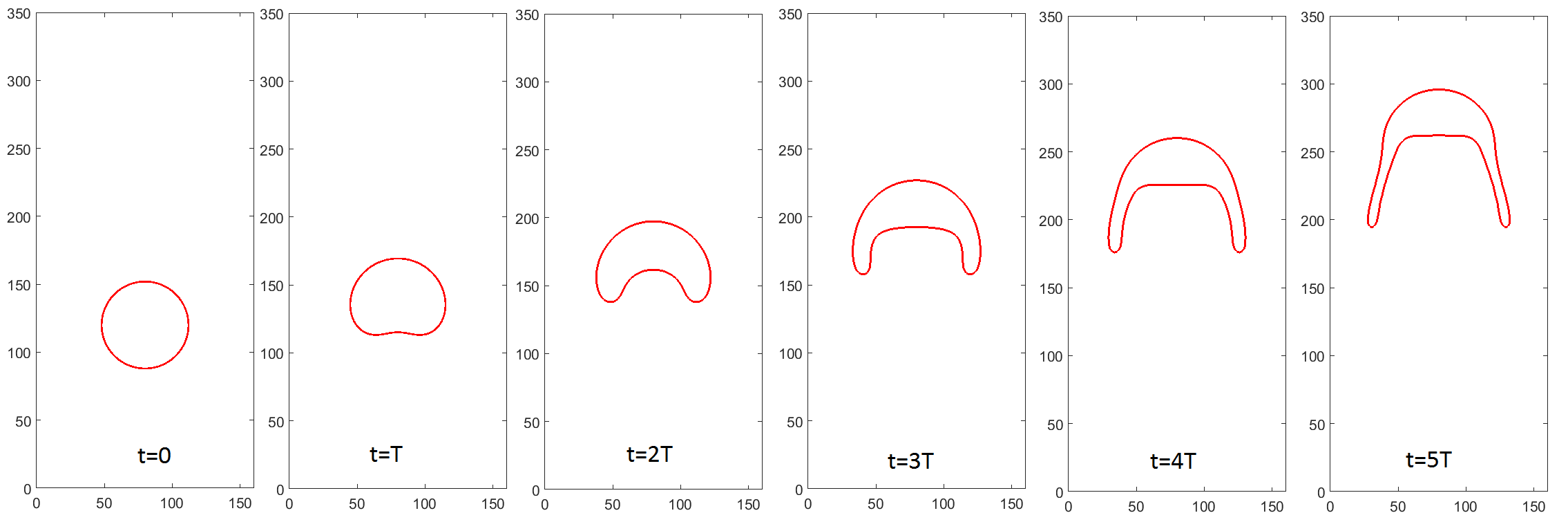}
         }
    \caption{Evolution of the interface of a buoyancy-driven rising bubble at $\mbox{Re}=35$ and (a) $\mbox{Eo}=10$, (b) $\mbox{Eo}=50$, (c) $\mbox{Eo}=125$.}
    \label{fig:rising3}
\end{figure}

\begin{figure}[htbp]
\centering
    \subfloat{
        \includegraphics[height=8cm, width=8cm]{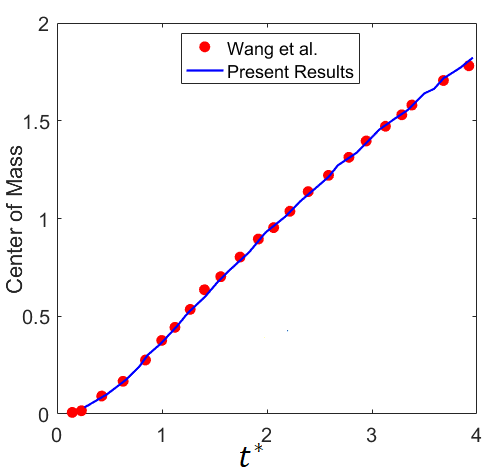}
         }
    \caption{Time history of the non-dimensional center of mass of a buoyancy-driven rising bubble at $\mbox{Re}=35$ and $\mbox{Eo}=125$.}
    \label{fig:centermass}
\end{figure}

By setting $\rho_A/\rho_B=1000$ and $\mu_A/\mu_B=100$ at a fixed Reynolds number $\mbox{Re}=35$, we performed buoyancy-driven bubble rise simulations at various values of the Eotvos numbers $\mbox{Eo}=10, 50$ and $125$ (as in Refs.~\cite{aland2012benchmark,wang2015mass}) using the cascaded LB methods. Figure~\ref{fig:rising3} presents the computed evolution of the interface of the rising bubble at these three values of $\mbox{Eo}$. When the role of the surface tension force is relatively significant in comparison with the other forces, as when the Eotvos number is low ($\mbox{Eo}=10$), the bubble undergoes smaller deformation that is initiated at its rear end, which then results in a flattening of that side as the bubble rises. For the intermediate case ($\mbox{Eo}=35$), the driving buoyancy force predominates the surface tension under the prevailing viscous force, resulting in a much larger deformation by stretching that leads to the formation of tails that elongates at later times. At even higher $\mbox{Eo}=125$, this process is more pronounced and the skirted shape accompanied by the pair of tails is further elongated and straightened. These computed shape variations with different $\mbox{Eo}$ at various time are very similar with the results based on other methods~\cite{aland2012benchmark,wang2015mass}.
Furthermore, in order to make a quantitative comparison, we then compute the vertical coordinate of the center of mass of the rising bubble as it undergoes shape changes using $y_c=\int_{\Omega_b}y dx/\int_{\Omega_b}1 dx$, where $\Omega_b$ represents the region occupied by the bubble, for the
case $\mbox{Re}=35$ and $\mbox{Eo}=125$. Figure~\ref{fig:centermass} shows the non-dimensional center of mass as a function of the non-dimensional time computed using the cascaded LB schemes against the reference numerical results from Ref.~\cite{wang2015mass}. It is evident that our approach
is in good quantitative agreement with the available numerical data for the temporal evolution of the bubble paths, thereby verifying its accuracy and robustness for this high density ratio two-phase flow problem.

\subsection{\label{sec:resultrisingbubble} Impact of a drop on a thin liquid layer}
As another case study, we consider an inertia-driven two-phase flow problem at a high density ratio, i.e., the impact of a circular drop on a thin layer of fluid and the study of its subsequent outcomes. Such impact dynamics of drops leads to a rich variety of outcomes depending on the
characteristic parameters representing the ratios of various attendant forces~\cite{yarin2006drop}. The computational set up considered for this example is described in Ref.~\cite{liang2018phase}. Both the drop and the thin layer are considered to be the of the same liquid of density $\rho_A$ and the ambient fluid is of density $\rho_B$. We consider a high density ratio $\rho_A/\rho_B=1000$ to represent the impact of a water drop surrounded by air. The computational domain is resolved with $501\times1501$ grid nodes, in which the liquid layer is discretized by $150$ grid nodes, while the drop radius $R$ is represented by $100$ mesh nodes. The interface thickness $W$ is set to be $5$. We impose periodic conditions on the two vertical boundaries, no-slip boundary condition on the bottom wall, and free-slip condition on the top boundary. The drop is set into downward motion by setting it with an initial impact velocity $U=0.05$. The dynamics and the impact outcomes of this problem is determined by the following non-dimensional parameters: the Reynolds number $\mbox{Re}=2\rho_AUR/\mu_A$ and the Weber number $\mbox{We}=2\rho_AU^2R/\sigma$, which represents the ratio of the inertial force to the surface tension force, in addition to the ratios of the fluid properties, and the timescale is given as $2R/U$. In our cascaded LB simulations, with the density ratio given above, we set $\mu_A/\mu_B=10$, the Weber number is fixed at $\mbox{We}=8000$, and consider two different values of the Reynolds numbers: $\mbox{Re}=20$ and $\mbox{Re}=100$.

\begin{figure}[htbp]
\centering
    \subfloat[\label{subfig-2:dummy}]{
        \includegraphics[height=12cm, width=7cm]{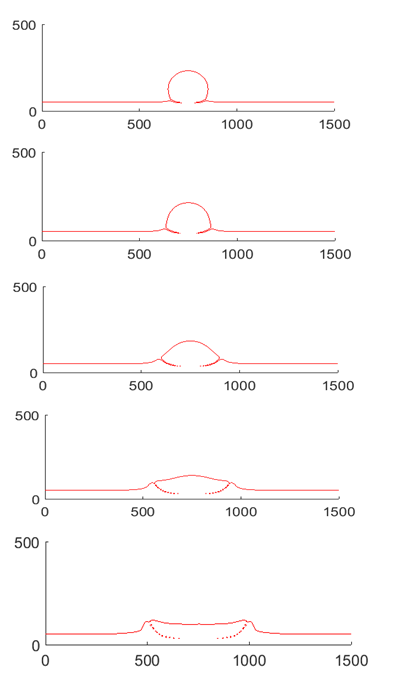}
         }
    \subfloat[\label{subfig-2:dummy}]{
        \includegraphics[height=12cm, width=7cm] {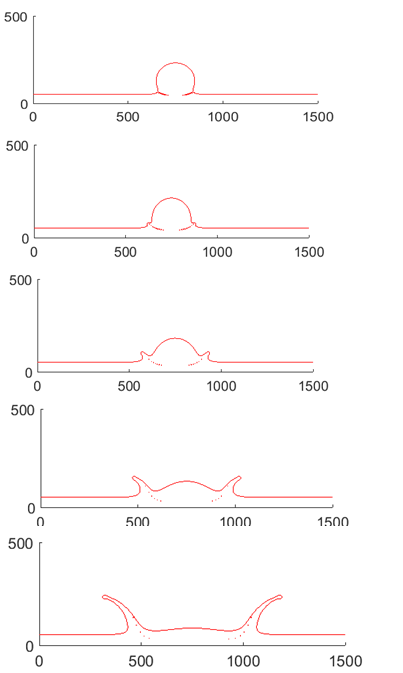}
         }
         \caption{Evolution of the splashing of a drop on a thin film at $\mbox{We}=8000$ and $\rho_A/\rho_B=1000$ for (a) $\mbox{Re}=20$ (b)$ \mbox{Re}=100$.}
    \label{fig:splashing}
\end{figure}

Figure~\ref{fig:splashing} presents the evolutions of interfaces at these two Reynolds numbers upon drop impact. At the lower $\mbox{Re}=20$, since the kinetic energy of the drop impact is relatively low, it merges with the liquid film, which is accompanied by the interfacial wave moving outwards. This results in the deposition of the drop as the outcome. On the other hand, as the $\mbox{Re}$ is increased to $100$, upon drop impact, the interface initially spreads outcomes, and then with the higher attendant kinetic energy, it leads an ejecta sheet formation. This, in turn,
spreads outwards by evolving into a splashing lamella that curls at its edges due to the competing surface tension and viscous frictional effects, leading to the splashing as the final outcome. These computed behaviors are consistent with other recent numerical results (e.g.,~\cite{liang2018phase}), which demonstrate the ability of the cascaded LB schemes to handle inertia-driven two-phase flows at high density ratios.

\subsection{\label{sec:resultMarangonistressdropmigration}Tangential surface tension gradient (Marangoni stress) effect on drop migration}
Variable surface tension effects arise in certain unique class of two-phase flows such as those involving thermo-capillary convection and surfactant-laden multiphase flows. For example, surfactants play an important role in numerous two-fluid dispersed systems where they strongly modulate
phenomena associated with droplets and bubbles by preferentially adsorbing on the interfaces with nonuniform distribution, which then lower the local surface tension and can induce additional fluid motion around interfaces via the tangential surface tension gradients or Marangoni stresses. Thus, the expression for the surface tension force $\bm{F}_s$ given earlier in Eq.~(\ref{eq:surfacetensiongeometric}) needs to be modified to account for surfactant effects. In this regard, we will adopt the geometric formulation presented in~\cite{yun2014new}. The smoothed surface tension formulation for surfacant-laden interfacial flows with a local surfactant concentration $\psi$ can be written as
\begin{equation}
\bm{F}_s=\underbrace{-\tilde{\kappa}(\psi)|\bm{\nabla} \phi|^2\left(\bm{\nabla}\cdot\bm{n}\right)\bm{n}}_{\mbox{Capillary force}}+\underbrace{|\bm{\nabla} \phi|^2\bm{\nabla}_s\tilde{\kappa}(\psi)}_{\mbox{Marangoni force}},\label{eq:surfacetensiongeometricsurfactants}
\end{equation}
where $\bm{\nabla}_s$ is the surface gradient operator given by $\bm{\nabla}_s\equiv \bm{\nabla}-\bm{n}(\bm{n}\cdot\bm{\nabla})$ or in index notation $\partial_{si}=(\delta_{ij}-n_in_j)\partial_j$, where $i, j\in (x,y)$. The first term on the RHS of Eq.~(\ref{eq:surfacetensiongeometricsurfactants}) represents the capillary force, where the lowering of the local surface tension by the presence of surfactant is accounted through the dependence of
the surface tension parameter $\tilde{\kappa}$ on $\psi$, i.e., $\tilde{\kappa}(\psi)$ (see below for details). The second term represents the effects of
the tangential gradients of the surface tension, or the Marangoni force, arising from the non-uniform concentration of the surfactant on the interface. The Cartesian components of the surface tension force for surfactant-laden interfaces can then be expressed as
\begin{subequations}
\begin{eqnarray}
F_{sx}&=&-\tilde{\kappa}(\psi)|\bm{\nabla} \phi|^2(\bm{\nabla}\cdot\bm{n})n_x+|\bm{\nabla} \phi|^2\left[(1-n_x^2)\partial_x\tilde{\kappa}(\psi)-n_xn_y\partial_y\tilde{\kappa}(\psi)\right],\label{eq:surfacetensiongeometricsurfactantsxcomp}\\
F_{sy}&=&-\tilde{\kappa}(\psi)|\bm{\nabla} \phi|^2(\bm{\nabla}\cdot\bm{n})n_y+|\bm{\nabla} \phi|^2\left[(1-n_y^2)\partial_y\tilde{\kappa}(\psi)-n_xn_y\partial_x\tilde{\kappa}(\psi)\right],\label{eq:surfacetensiongeometricsurfactantsycomp}
\end{eqnarray}
\end{subequations}
where $n_x$ and $n_y$ are the components of the interfacial unit normal $\bm{n}=(n_x,n_y)=\bm{\nabla}\phi/|\bm{\nabla}\phi|$. Such a geometric strategy enhances flexibility as the effect of surfactant on the surface tension force is naturally tunable with an appropriate choice of the interfacial equation of the state. In this work, the interface equation of state to represent the influence of the surfactant on (lowering) the local surface tension is given by the following non-linear dependence based on the Langmuir isotherm, i.e., $\sigma(\psi)=\sigma_0\left[1+\beta \ln (1-\psi)\right]$, or, equivalently
\begin{equation}
\tilde{\kappa}(\psi)=\tilde{\kappa}_0\left[1+\beta \ln (1-\psi)\right],\label{eq:Langmuirisotherm}
\end{equation}
where $\beta$ is the Gibbs elasticity number that parameterizes the sensitivity of the surface tension to the local surfactant concentration, and $\sigma_0$ and $\tilde{\kappa}_0$ correspond to those for the clean interfaces, i.e., without the presence of surfactant.

In general, the above formulation would require computing the evolution of the surfactant concentration $\psi$. This can be accomplished by means of
a phase-field model for surfactant dynamics and an additional cascaded LB scheme for its solution procedure~\cite{Hajabdollahiphdthesis}. However, here
the focus will be on validating the implementation of the surface tension force, i.e., Eqs.~(\ref{eq:surfacetensiongeometricsurfactantsxcomp}) and (\ref{eq:surfacetensiongeometricsurfactantsycomp}), and in particular the Marangoni force, in our formulation for an \emph{imposed} surfactant concentration profile for which an analytical solution for the motion of the dispersed phase is available for making a comparison. In this regard, we consider the classical Young's problem of thermocapillary migration of a drop~\cite{young1959motion,subramanian2001motion} and recast into the equivalent surfactant concentration gradient driven problem. According to this problem, a neutrally-buoyant drop of fluid $A$ with diameter $D$ solely under an imposed linear surfactant concentration profile $\psi(y)=a+G_\Gamma y$ (i.e., $G_\Gamma$ being the constant gradient of the surfactant concentration field and $y$ is the vertical coordinate) will self-propel in the ambient fluid $B$  and its terminal migration velocity under the assumption of creeping flow has the following analytical solution:
\begin{equation*}
V_{\Gamma}=-\frac{\sigma_\Gamma G_\Gamma D}{6\mu_B+9\mu_A},
\end{equation*}
where $\sigma_\Gamma$ is the sensitivity of the surface tension with the surfactant concentration, which, according to the linearized form of the Langmuir's isotherm for dilute surfactant concentration, can be expressed as $\sigma_\Gamma\equiv \partial\sigma/\partial\psi=-\sigma_0\beta$. $\mu_A$ and $\mu_B$ are the respective dynamic viscosities.

\begin{figure}[htbp]
\centering
    \subfloat[\label{subfig-2:dummy}]{
        \includegraphics[height=8cm, width=8cm]{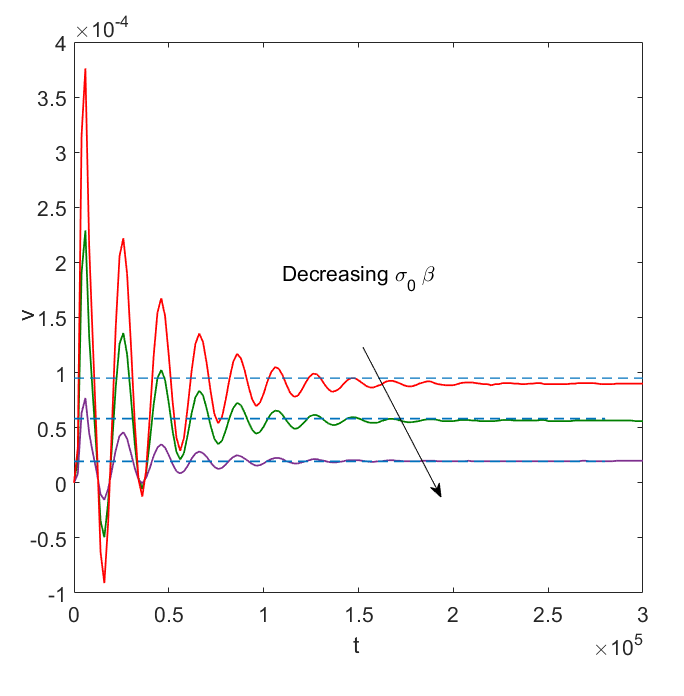}
         }
    \caption{Comparison of computed drop migration velocity under imposed constant surfactant concentration gradient in the simulation of Young's problem (solid lines) with the analytical solution for the terminal velocity (dashed lines) for surface tension sensitivities $\sigma_0\beta=0.0048, \sigma_0\beta=0.0146$ and $\sigma_0\beta=0.0244$.}
    \label{fig:final11}
\end{figure}

\begin{figure}[htbp]
\centering
    \subfloat[\label{subfig-2:dummy}]{
        \includegraphics[height=9cm, width=9cm]{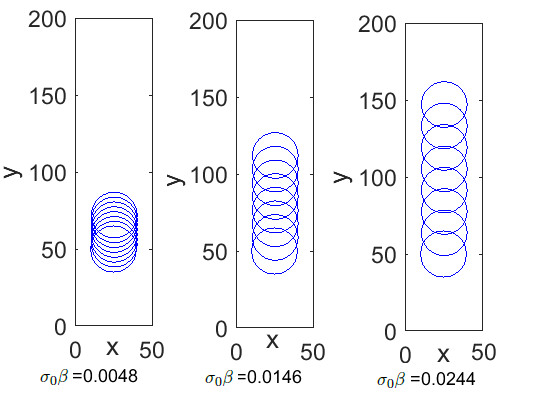}
         }
    \caption{Snapshots of the evolution of a migrating drop under imposed constant surfactant concentration gradient in the simulation of Young's problem for surface tension sensitivities $\sigma_0\beta=0.0048, \sigma_0\beta=0.0146 $ and $\sigma_0\beta=0.0244$.}
    \label{fig:moving}
\end{figure}

We consider a drop with diameter $D=30$ initially located near the bottom of a rectangular domain resolved with $51\times 201$ grid nodes. Periodic boundary conditions along the two vertical sides and no-slip boundary conditions along the two horizontal sides are imposed. By using a density ratio of unity, we consider the same dynamic viscosities in both the fluids by setting the kinematic viscosities as $\nu_A=\nu_B=0.05$. Furthermore, we impose a linear variation of the surfactant concentration along the vertical direction by setting its slope $G_\Gamma=9.95\times 10^{-5}$. Figure~\ref{fig:final11} shows the computed the drop migration velocities for three different surface tension sensitivities $\sigma_0\beta=0.0048, \sigma_0\beta=0.0146 $ and $\sigma_0\beta=0.0244$ and their comparisons against the available analytical solution for the terminal velocity. It is evident that after the initial transients, the computed migration velocities in the long time limit are in good agreement with the analytical terminal velocity.
In addition, some snapshots of the evolution of a migrating drop for all the above three cases are presented in Fig.~\ref{fig:moving}. As it can be seen, the drop self-propels under non-uniform surface tension (i.e., Marangoni force) arising due to an imposed constant concentration gradient without any smearing effects to the shape of the drop. Thus, the above numerical simulation results validate our implementation for handling variable surface tension effects. A more general case of the coupled evolution of the surfactant concentration field, two-fluid motion and interface advection via unified cascaded LB formulations~\cite{Hajabdollahiphdthesis}, and its application for studying the physics of surfactant-laden two-fluid systems are subjects of future investigations.

\section{\label{sec:stabilitycomparison}Comparative study of numerical stability of different collision models}
\begin{figure}[htbp]
\centering
    \subfloat[\label{subfig-2:dummy}]{
        \includegraphics[height=8cm, width=8cm]{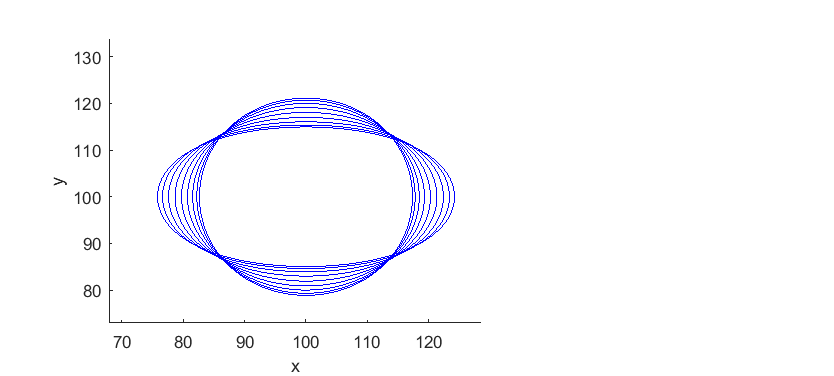}
         }
                                     \caption{Evolution of the interface of an oscillating liquid cylinder starting from an initial elliptic shape configuration with semi-major axis $a=25$ and semi-minor axis $b=15$; surface tension parameter $\tilde{\kappa} = 0.1$, kinematic viscosity $\nu_A =\nu_B= 0.01$ and density ratio $\rho_A/\rho_B=100$.}
    \label{fig:bubble}
\end{figure}

\begin{figure}[htbp]
\centering
    \subfloat[\label{subfig-2:dummy}]{
        \includegraphics[height=8cm, width=8cm]{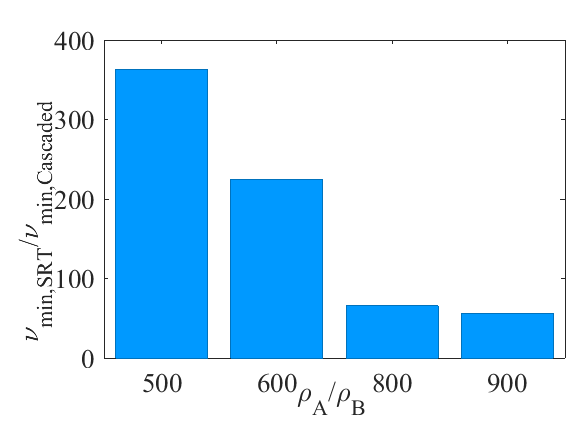}
         }
                                     \caption{Comparison of the ratios of the minimum achievable viscosities for single relaxation time (SRT) and cascaded LB formulations allowing numerically stable simulations of an oscillating liquid cylinder with surface tension parameter $\tilde{\kappa} = 0.01$ at different density ratios.}
    \label{fig:u1}
\end{figure}

Generally, it is known that the LB methods can be susceptible to numerical instabilities as the kinematic viscosity of the fluid being simulated is significantly lowered, which is strongly influenced by the type of collision model used. We will now assess the robustness of our cascaded LB formulation in achieving relatively low fluid kinematic viscosities, when compared to a single relaxation time (SRT) formulation for a two-fluid case study involving capillary oscillations of a liquid cylinder in another ambient lighter fluid. Prior studies have considered such a configuration in assessing the numerical stability of the LB schemes for two-phase flows~\cite{Premnath2007,mccracken2005multiple}. The SRT formulation for two-phase flows used for comparison is based on one SRT LB solver obtained as a discretization of the MCBE for two-phase fluid motion and another SRT LB scheme for capturing interfacial dynamics represented by the conservative ACE. We consider a periodic domain of resolution $200 \times 200$ in which a liquid cylinder of
density $\rho_A$ is placed in another lighter ambient fluid of density $\rho_B$, where $\nu_A=\nu_B$ for simplicity, undergoes free oscillations. The oscillations are initiated from an initially elliptic configuration of the cylinder (semi-major axis $a=25$ and semi-minor axis $b=15$) via the capillary effects on its interface. Figure~\ref{fig:bubble} shows a typical example of the evolution of the interface of the liquid cylinder undergoing free oscillations.
Now, employing each of the two collision models, for the above initial geometric configuration of the liquid cylinder with surface tension parameter $\tilde{\kappa}=0.01$, and for four sets of values of the density ratios $\rho_A/\rho_B=500, 600, 800$ and $900$, the kinematic viscosity of the fluids $\nu_A=\nu_B$ are gradually reduced till the simulations becomes unstable. Figure~\ref{fig:u1} reports the ratios of the minimum achievable viscosities for SRT and cascaded LB formulations that allow stable simulations for the above values of density ratios.
It is evident that dramatic improvements in numerical stability, by over one or two orders of magnitude, is achieved by the cascaded LB schemes when compared to the SRT LB schemes for this two-fluid case study. For example, even at high density ratio of $900$, the lowest viscosity achieved by the cascaded LB schemes is smaller by a factor of over 55, when compared to that attained using the SRT LB schemes, and such factors are significantly higher at more moderate density ratios. These numerical stability improvements associated with using the cascaded LB formulations for two-phase flow simulations are consistent with the findings of previous studies on single-phase flows (e.g.,~\cite{ning2016numerical,hajabdollahi2019cascaded}).

\section{\label{sec:summaryandconclusions}Summary and Conclusions}
In this paper, we discussed new cascaded LB formulations based on central moments and multiple relaxation times for computation of two-phase, incompressible flows at high density ratios. Using the modified continuous Boltzmann equation (MCBE) for two-phase flows, which involves a kinetic transformation to handle numerical stiffness at high density gradients, as a starting point, a cascaded LB scheme for the solution of the incompressible two-phase fluid motion directly in terms of the pressure and velocity fields is constructed. This involves the representation of the collision step via the relaxation of various central moments to their equilibria that are obtained by matching the corresponding continuous central moments of the modified Maxwell distribution expressed in terms of the pressure field. In addition, a consistent forcing scheme to handle the surface tension and body forces, as well as the net gradient pressure force, whose effects on the changes in various moments are different, is constructed. In order to capture the interfacial dynamics, another cascaded LB method that solves the phase field based conservative Allen-Cahn equation (ACE), which evolves interfaces by advection due to fluid motion under competing effects of diffusion and sharpening terms, is developed. This is achieved by a modification of first order central moments of the corresponding equilibrium distribution function via the addition of the interface sharpening term. Simulations of a variety of benchmark problems, including the equilibrium of a static drop, Rayleigh-Taylor instability, falling drop under gravity, buoyancy-driven rising bubble, drop impact on a thin liquid layer, validated the ability of the cascaded LB schemes to reproduce complex two-phase interfacial flows at high density ratios with good accuracy.
In addition, we showed that our formulation can be extended to handle variable surface tension effects by its validation for the simulation of the migration of neutrally buoyant drop under tangential surface tension gradients. Furthermore, dramatic improvements in numerical stability in reaching relatively low viscosities in two-phase systems with the use of cascaded LB approach when compared to a single relaxation time formulation is demonstrated. Thus, the cascaded LB methods for coupled solution of the fluid motion and interfacial dynamics, based on the MCBE and conservative ACE, are accurate and robust for two-phase flow simulations with high contrasts in fluid properties and with tunable surface tension effects. Future work includes an extension of this formulation to three-dimensions for simulation of surfactant-laden multiphase flows.

\section*{Acknowledgements}
The authors would like to acknowledge the support of the US National Science Foundation (NSF) under Grant CBET-1705630. This work was presented at the 71st Annual Meeting of the APS Division of Fluid Dynamics (DFD), Atlanta, Georgia, Nov. 2018 (\url{http://meetings.aps.org/link/BAPS.2018.DFD.L31.7}) with travel support from NSF. The first author's Ph.D. dissertation~\cite{Hajabdollahiphdthesis} is based, in part, on the research contribution presented in this work and she wishes to thank the dissertation committee members for the various constructive comments and suggestions.


\begin{thebibliography}{10}
\expandafter\ifx\csname url\endcsname\relax
  \def\url#1{\texttt{#1}}\fi
\expandafter\ifx\csname urlprefix\endcsname\relax\def\urlprefix{URL }\fi
\expandafter\ifx\csname href\endcsname\relax
  \def\href#1#2{#2} \def\path#1{#1}\fi

\bibitem{schwarzkopf2011multiphase}
J.~D. Schwarzkopf, M.~Sommerfeld, C.~T. Crowe, Y.~Tsuji, Multiphase flows with
  droplets and particles, CRC press, 2011.

\bibitem{scardovelli1999direct}
R.~Scardovelli, S.~Zaleski, Direct numerical simulation of free-surface and
  interfacial flow, Annual review of fluid mechanics 31~(1) (1999) 567--603.

\bibitem{tryggvason2001front}
G.~Tryggvason, B.~Bunner, A.~Esmaeeli, D.~Juric, N.~Al-Rawahi, W.~Tauber,
  J.~Han, S.~Nas, Y.-J. Jan, A front-tracking method for the computations of
  multiphase flow, Journal of Computational Physics 169~(2) (2001) 708--759.

\bibitem{osher2006level}
S.~Osher, R.~Fedkiw, Level set methods and dynamic implicit surfaces, Vol. 153,
  Springer Science \& Business Media, 2006.

\bibitem{he1997theory}
X.~He, L.-S. Luo, Theory of the lattice {B}oltzmann method: From the
  {B}oltzmann equation to the lattice {B}oltzmann equation, Phys. Rev. E 56~(6)
  (1997) 6811.

\bibitem{Chen1998}
S.~Chen, G.~Doolen, Lattice {B}oltzmann method for fluid flows, Ann. Rev. Fluid
  Mech. 30 (1998) 329--364.

\bibitem{Succi2001}
S.~Succi, The Lattice Boltzmann equation for fluid dynamics and beyond, Oxford
  University Press., New York, 2001.

\bibitem{Aidun2010}
C.~Aidun, J.~Clausen, Lattice-{B}oltzmann method for complex flows, Annu. Rev.
  Fluid Mech. 42 (2010) 439--472.

\bibitem{Kruger2016}
T.~Kruger, H.~Kusumaatmaja, A.~Kuzmin, O.~Shardt, G.~Silva, E.~Viggen, The
  Lattice {B}oltzmann Method - Principles and Practice, Springer, 2016.

\bibitem{gunstensen1991lattice}
A.~K. Gunstensen, D.~H. Rothman, S.~Zaleski, G.~Zanetti, Lattice {B}oltzmann
  model of immiscible fluids, Physical Review A 43~(8) (1991) 4320.

\bibitem{grunau1993lattice}
D.~Grunau, S.~Chen, K.~Eggert, A lattice {B}oltzmann model for multiphase fluid
  flows, Physics of Fluids A: Fluid Dynamics 5~(10) (1993) 2557--2562.

\bibitem{shan1993lattice}
X.~Shan, H.~Chen, Lattice {B}oltzmann model for simulating flows with multiple
  phases and components, Physical Review E 47~(3) (1993) 1815.

\bibitem{swift1996lattice}
M.~R. Swift, E.~Orlandini, W.~Osborn, J.~Yeomans, Lattice {B}oltzmann
  simulations of liquid-gas and binary fluid systems, Physical Review E 54~(5)
  (1996) 5041.

\bibitem{Luo2000}
L.-S. Luo, Theory of the lattice {B}oltzmann method: Lattice {B}oltzmann models
  for nonideal gases, Phys. Rev. E. 62 (2000) 4982--4996.

\bibitem{He2002}
X.~He, G.~Doolen, Thermodynamic foundations of kinetic theory and lattice
  {B}oltzmann models for multiphase flows, J. Stat. Phys. 107 (2002)
  1572--4996.

\bibitem{wagner2006thermodynamic}
A.~Wagner, Thermodynamic consistency of liquid-gas lattice {B}oltzmann
  simulations, Phys. Rev. E 74~(5) (2006) 056703.

\bibitem{he1999lattice}
X.~He, S.~Chen, R.~Zhang, A lattice {B}oltzmann scheme for incompressible
  multiphase flow and its application in simulation of rayleigh--taylor
  instability, Journal of Computational Physics 152~(2) (1999) 642--663.

\bibitem{lee2005stable}
T.~Lee, C.-L. Lin, A stable discretization of the lattice {B}oltzmann equation
  for simulation of incompressible two-phase flows at high density ratio,
  Journal of Computational Physics 206~(1) (2005) 16--47.

\bibitem{anderson1998diffuse}
D.~M. Anderson, G.~B. McFadden, A.~A. Wheeler, Diffuse-interface methods in
  fluid mechanics, Annual review of fluid mechanics 30~(1) (1998) 139--165.

\bibitem{jacqmin1999calculation}
D.~Jacqmin, Calculation of two-phase {N}avier--{S}tokes flows using phase-field
  modeling, Journal of Computational Physics 155~(1) (1999) 96--127.

\bibitem{ding2007diffuse}
H.~Ding, P.~D. Spelt, C.~Shu, Diffuse interface model for incompressible
  two-phase flows with large density ratios, Journal of Computational Physics
  226~(2) (2007) 2078--2095.

\bibitem{kim2012phase}
J.~Kim, Phase-field models for multi-component fluid flows, Communications in
  Computational Physics 12~(3) (2012) 613--661.

\bibitem{cahn1958free}
J.~W. Cahn, J.~E. Hilliard, Free energy of a nonuniform system. i. interfacial
  free energy, The Journal of chemical physics 28~(2) (1958) 258--267.

\bibitem{zheng2006lattice}
H.~Zheng, C.~Shu, Y.-T. Chew, A lattice {B}oltzmann model for multiphase flows
  with large density ratio, Journal of Computational Physics 218~(1) (2006)
  353--371.

\bibitem{fakhari2010phase}
A.~Fakhari, M.~H. Rahimian, Phase-field modeling by the method of lattice
  {B}oltzmann equations, Physical Review E 81~(3) (2010) 036707.

\bibitem{zu2013phase}
Y.~Zu, S.~He, Phase-field-based lattice {B}oltzmann model for incompressible
  binary fluid systems with density and viscosity contrasts, Physical Review E
  87~(4) (2013) 043301.

\bibitem{liang2014phase}
H.~Liang, B.~Shi, Z.~Guo, Z.~Chai, Phase-field-based multiple-relaxation-time
  lattice {B}oltzmann model for incompressible multiphase flows, Physical
  Review E 89~(5) (2014) 053320.

\bibitem{allen1976mechanisms}
S.~M. Allen, J.~W. Cahn, Mechanisms of phase transformations within the
  miscibility gap of {F}e-rich {F}e-{A}l alloys, Acta Metallurgica 24~(5)
  (1976) 425--437.

\bibitem{folch1999phase}
R.~Folch, J.~Casademunt, A.~Hern{\'a}ndez-Machado, L.~Ramirez-Piscina,
  Phase-field model for {H}ele-{S}haw flows with arbitrary viscosity contrast.
  i. theoretical approach, Physical Review E 60~(2) (1999) 1724.

\bibitem{sun2007sharp}
Y.~Sun, C.~Beckermann, Sharp interface tracking using the phase-field equation,
  Journal of Computational Physics 220~(2) (2007) 626--653.

\bibitem{chiu2011conservative}
P.-H. Chiu, Y.-T. Lin, A conservative phase field method for solving
  incompressible two-phase flows, Journal of Computational Physics 230~(1)
  (2011) 185--204.

\bibitem{olsson2005conservative}
E.~Olsson, G.~Kreiss, A conservative level set method for two phase flow,
  Journal of computational physics 210~(1) (2005) 225--246.

\bibitem{geier2015conservative}
M.~Geier, A.~Fakhari, T.~Lee, Conservative phase-field lattice {B}oltzmann
  model for interface tracking equation, Physical Review E 91~(6) (2015)
  063309.

\bibitem{ren2016improved}
F.~Ren, B.~Song, M.~C. Sukop, H.~Hu, Improved lattice {B}oltzmann modeling of
  binary flow based on the conservative {A}llen-{C}ahn equation, Physical
  Review E 94~(2) (2016) 023311.

\bibitem{Qian1992}
Y.~Qian, D.~D. Humieres, P.~Lallemand, Lattice {BGK} models for
  {N}avier-{S}tokes equation, Europhysics Letters. 17 (1992) 479--484.

\bibitem{dHumieres2002}
D.~d'Humieres, I.~Ginzburg, M.~Krafczyk, P.~Lallemand, L.-S. Luo.,
  Multiple-relaxation-time lattice {B}oltzmann models in three dimensions,
  Phil. Trans. R. Soc. Lond. A. 360 (2002) 437--451.

\bibitem{geier2006cascaded}
M.~Geier, A.~Greiner, J.~G. Korvink, Cascaded digital lattice {B}oltzmann
  automata for high reynolds number flow, Physical Review E 73~(6) (2006)
  066705.

\bibitem{Asinari2008}
P.~Asinari, Generalized local equilibrium in the cascaded lattice {B}oltzmann
  method, Phys. Rev. E 78 (2008) 016701.

\bibitem{premnath2009incorporating}
K.~N. Premnath, S.~Banerjee, Incorporating forcing terms in cascaded lattice
  {B}oltzmann approach by method of central moments, Physical Review E 80~(3)
  (2009) 036702.

\bibitem{Geier2015}
M.~Geier, M.~Schonherr, A.~Pasquali, M.~Krafczyk, The cumulant lattice
  {B}oltzmann equation in three dimensions: Theory and validation, Comp. Math.
  Appl. 704 (2015) 507--547.

\bibitem{ning2016numerical}
Y.~Ning, K.~N. Premnath, D.~V. Patil, Numerical study of the properties of the
  central moment lattice {B}oltzmann method, International Journal for
  Numerical Methods in Fluids 82~(2) (2016) 59--90.

\bibitem{lycett2016cascaded}
D.~Lycett-Brown, K.~H. Luo, Cascaded lattice {B}oltzmann method with improved
  forcing scheme for large-density-ratio multiphase flow at high {R}eynolds and
  {W}eber numbers, Physical Review E 94~(5) (2016) 053313.

\bibitem{de2017non}
A.~De~Rosis, Non-orthogonal central moments relaxing to a discrete equilibrium:
  A d2q9 lattice {B}oltzmann model, EPL (Europhysics Letters) 116~(4) (2017)
  44003.

\bibitem{fei2018cascaded}
L.~Fei, K.~H. Luo, Cascaded lattice {B}oltzmann method for incompressible
  thermal flows with heat sources and general thermal boundary conditions,
  Computers \& Fluids 165 (2018) 89--95.

\bibitem{hajabdollahi2018sym}
F.~Hajabdollahi, K.~N. Premnath, Symmetrized operator split schemes for force
  and source modeling in cascaded lattice boltzmann methods for flow and scalar
  transport, Physical Review E 97~(6) (2018) 063303.

\bibitem{elseid2018cascaded}
F.~M. Elseid, S.~W. Welch, K.~N. Premnath, A cascaded lattice {B}oltzmann model
  for thermal convective flows with local heat sources, International Journal
  of Heat and Fluid Flow 70 (2018) 279--298.

\bibitem{fei2018modeling}
L.~Fei, K.~H. Luo, C.~Lin, Q.~Li, Modeling incompressible thermal flows using a
  central-moments-based lattice {B}oltzmann method, International Journal of
  Heat and Mass Transfer 120 (2018) 624--634.

\bibitem{hajabdollahi2018central}
F.~Hajabdollahi, K.~N. Premnath, Central moments-based cascaded lattice
  {B}oltzmann method for thermal convective flows in three-dimensions,
  International Journal of Heat and Mass Transfer 120 (2018) 838--850.

\bibitem{safari2018lattice}
H.~Safari, M.~Krafczyk, M.~Geier, A lattice {B}oltzmann model for thermal
  compressible flows at low {M}ach numbers beyond the boussinesq approximation,
  Computers \& Fluids.

\bibitem{hajabdollahi2019cascaded}
F.~Hajabdollahi, K.~N. Premnath, S.~W. Welch, Cascaded lattice {B}oltzmann
  method based on central moments for axisymmetric thermal flows including
  swirling effects, International Journal of Heat and Mass Transfer 128 (2019)
  999--1016.

\bibitem{hajabdollahiAPSDFD2018}
F.~Hajabdollahi, K.~Premnath, S.~W. Welch,
  \href{http://meetings.aps.org/link/BAPS.2018.DFD.L31.7}{Cascaded lattice
  boltzmann method for phase-field modeling of incompressible multiphase
  flows}, in: Bulletin of the American Physical Society of the 71st Annual
  Meeting of the APS Division of Fluid Dynamics (DFD), Atlanta, Georgia, 2018.
\newline\urlprefix\url{http://meetings.aps.org/link/BAPS.2018.DFD.L31.7}

\bibitem{Hajabdollahiphdthesis}
F.~Hajabdollahi, Cascaded lattice {B}oltzmann methods based on central moments
  for thermal convection, multiphase flows and complex fluids, Ph.D. thesis,
  University of Colorado Denver, Denver, CO (March 2019).

\bibitem{popinet2018numerical}
S.~Popinet, Numerical models of surface tension, Annual Review of Fluid
  Mechanics 50 (2018) 49--75.

\bibitem{kim2005continuous}
J.~Kim, A continuous surface tension force formulation for diffuse-interface
  models, Journal of Computational Physics 204~(2) (2005) 784--804.

\bibitem{Premnath2007}
K.~N. Premnath, J.~Abraham, Three-dimensional multi-relaxation time ({MRT})
  lattice-{B}oltzmann models for multiphase flow, J. Comput. Phy 224 (2007)
  539--559.

\bibitem{rudman1997volume}
M.~Rudman, Volume-tracking methods for interfacial flow calculations,
  International Journal for Numerical Methods in Fluids 24~(7) (1997) 671--691.

\bibitem{wang2015mass}
Y.~Wang, C.~Shu, J.~Shao, J.~Wu, X.~Niu, A mass-conserved diffuse interface
  method and its application for incompressible multiphase flows with large
  density ratio, Journal of Computational Physics 290 (2015) 336--351.

\bibitem{li2012additional}
Q.~Li, K.~Luo, Y.~Gao, Y.~He, Additional interfacial force in lattice
  {B}oltzmann models for incompressible multiphase flows, Physical Review E
  85~(2) (2012) 026704.

\bibitem{fakhari2009simulation}
A.~Fakhari, M.~H. Rahimian, Simulation of falling droplet by the lattice
  {B}oltzmann method, Communications in Nonlinear Science and Numerical
  Simulation 14~(7) (2009) 3046--3055.

\bibitem{hysing2009quantitative}
S.-R. Hysing, S.~Turek, D.~Kuzmin, N.~Parolini, E.~Burman, S.~Ganesan,
  L.~Tobiska, Quantitative benchmark computations of two-dimensional bubble
  dynamics, International Journal for Numerical Methods in Fluids 60~(11)
  (2009) 1259--1288.

\bibitem{aland2012benchmark}
S.~Aland, A.~Voigt, Benchmark computations of diffuse interface models for
  two-dimensional bubble dynamics, International Journal for Numerical Methods
  in Fluids 69~(3) (2012) 747--761.

\bibitem{clift1978bubbles}
R.~Clift, J.~R. Grace, M.~E. Weber, Bubbles, drops, and particles, Academic
  Press, 1978.

\bibitem{yarin2006drop}
A.~L. Yarin, Drop impact dynamics: splashing, spreading, receding, bouncing…,
  Annu. Rev. Fluid Mech. 38 (2006) 159--192.

\bibitem{liang2018phase}
H.~Liang, J.~Xu, J.~Chen, H.~Wang, Z.~Chai, B.~Shi, Phase-field-based lattice
  {B}oltzmann modeling of large-density-ratio two-phase flows, Physical Review
  E 97~(3) (2018) 033309.

\bibitem{yun2014new}
A.~Yun, Y.~Li, J.~Kim, A new phase-field model for a water--oil-surfactant
  system, Applied Mathematics and Computation 229 (2014) 422--432.

\bibitem{young1959motion}
N.~Young, J.~Goldstein, M.~J. Block, The motion of bubbles in a vertical
  temperature gradient, Journal of Fluid Mechanics 6~(3) (1959) 350--356.

\bibitem{subramanian2001motion}
R.~S. Subramanian, R.~Balasubramaniam, The motion of bubbles and drops in
  reduced gravity, Cambridge University Press, 2001.

\bibitem{mccracken2005multiple}
M.~E. McCracken, J.~Abraham, Multiple-relaxation-time lattice-boltzmann model
  for multiphase flow, Physical Review E 71~(3) (2005) 036701.

\end{thebibliography}

\end{document}